\documentclass[preprint2]{aastex}

\slugcomment{manuscript for AJ, version 120720}

\shorttitle{UCAC4 release paper}
\shortauthors{Zacharias et al.}

\begin{document}


\title{The Fourth US Naval Observatory CCD Astrograph Catalog (UCAC4)}



\author{N. Zacharias$^1$,
        C.T. Finch$^1$,
        T.M. Girard$^2$,
        A. Henden$^3$, \\
        J.L. Bartlett$^1$,
        D.G. Monet$^4$,
        M.I. Zacharias$^5$}

\email{nz@usno.navy.mil}

\affil{$^1$U.S.~Naval Observatory, 3450 Mass.Ave.~NW Washington, DC 20392; \\
       $^2$Yale University, P.O. Box 208101, New Haven, CT 06520;\\
       $^3$director, AAVSO, Boston, MA;\\
       $^4$Naval Observatory Flagstaff Station (NOFS);\\
       $^5$contractor with USNO}


\begin{abstract}
The fourth United States Naval Observatory (USNO) CCD Astrograph Catalog, 
UCAC4 was released in August 2012 (double-sided DVD and CDS data center
Vizier catalog I/322).
It is the final release in this series and contains over 113 million
objects; over 105 million of them with proper motions.
UCAC4 is an updated version of UCAC3 with about the same number
of stars also covering all-sky.  Bugs were fixed, 
Schmidt plate survey data were avoided, and precise 5-band
photometry were added for about half the stars.
Astrograph observations have been supplemented for bright stars
by FK6, Hipparcos and Tycho-2 data to compile a UCAC4 star catalog 
complete from the brightest stars to about magnitude R = 16.
Epoch 1998 to 2004 positions are obtained from observations with the
20 cm aperture USNO Astrograph's ``red lens", equipped with a 4k by 4k CCD.
Mean positions and proper motions are derived by combining these 
observations with over 140 ground- and space-based catalogs, including 
Hipparcos/Tycho and the AC2000.2, as well as unpublished 
measures of over 5000 plates from other astrographs.  
For most of the faint stars 
in the Southern Hemisphere the first epoch plates from 
the Southern Proper Motion (SPM) program form the basis for 
proper motions, while the Northern Proper Motion (NPM) 1st epoch plates 
serve the same purpose for the rest of the sky. 
These data are supplemented by 2MASS near-IR photometry for about
110 million stars and 5-band (B,V,g,r,i) APASS data for over 
51 million stars.
Thus the published UCAC4, as were UCAC3 and UCAC2, is a compiled catalog 
with the UCAC observational program being a major component.
The positional accuracy of stars in UCAC4 at mean epoch is about 
15 to 100 mas per coordinate, depending on magnitude, while the formal
errors in proper motions range from about 1 to 10 mas/yr depending on 
magnitude and observing history.  Systematic errors in proper motions
are estimated to be about 1 to 4 mas/yr.
\end{abstract}


\section{Introduction}

The fourth United States Naval Observatory (USNO) CCD Astrograph Catalog,
UCAC4 is an all-sky, astrometric catalog of 113,780,093 objects
complete to about magnitude 16 (Fig.~1) in the instrumental system,
which is close to R magnitudes.  It contains accurate 
positions and proper motions on the International Celestial Reference 
System (ICRS) at a mean epoch around 2000, observed magnitudes, 
2-Micro All-Sky Survey (2MASS) magnitudes for about 110 million stars,
and AAVSO Photometric All-Sky Survey (APASS) 5-band photometry for over 
51 million stars.  
The published UCAC4 data are a compiled catalog of positions and
proper motions, based on final reductions of the UCAC observational
catalog of positions and many other astrometric catalogs.

The USNO operated the 8-inch (0.2 m) Twin
Astrograph from 1998 to 2004 for an all-sky astrometric survey.
About 2/3 of the sky was observed from the Cerro Tololo Inter-American
Observatory (CTIO) while the rest of the northern sky was observed
from the Naval Observatory Flagstaff Station (NOFS).
A 4k by 4k CCD with 9 $\mu$m pixel size was used at the ``red lens"
of the astrograph in a single bandpass (579 to 643 nm) with just over 
1 square degree field of view.
A 2-fold overlap pattern of fields span the entire sky.
Each field was observed with a long (about 125 sec) and a short
(about 25 sec) exposure. Thus each star should appear on at least
2 different CCD exposures, and stars in the mid-magnitude range
(about 10 to 14) should have 4 images.

UCAC4 is an incremental update of the large amount of work performed
earlier to arrive at the UCAC1 \citep{ucac1}), UCAC2 \citep{ucac2},
and UCAC3 \citep{ucac3r} data releases.
The positional precision of the CCD observations and the mean
UCAC4 catalog positions are very similar to those of UCAC3;
see statistics presented in \citep{ucac3r}.
The most significant improvement of UCAC4 over UCAC3 is the lower 
level of systematic errors in the proper motions north of the
Southern Proper Motion (SPM) area ($\delta \ge -20^{\circ}$).
All-sky color plots of proper motions and number of stars per sky
area are provided on the UCAC4 release DVD.
Table 1 gives some general statistics about UCAC4 data
(see also the ``readme" file of the public release).

As with UCAC3, the UCAC4 release catalog is based on all
applicable, regular survey field observations, excluding the CCD
exposures taken on extragalactic link fields and most calibration
fields.  Observations of minor planets have been extracted and will
be published separately.
The released UCAC4 is a compiled catalog, similar to UCAC3, and UCAC2.
However, we plan to make individual epoch CCD observations available 
in the future if desired by the community (please contact lead author).
Separate papers about new double stars found in the UCAC observational
catalog and the extragalactic link to International Celestial
Reference Frame (ICRF) quasars are in preparation.

\section{UCAC4 versus UCAC3}

The main differences between UCAC4 and UCAC3 data are the following:

\begin{description}

\item[Bug fixes:] add missing stars, remove multiple entries, apply 
  magnitude equation (CTE) corrections properly for all exposure times.

\item [NPM data:] are used to derive proper motions of faint stars north
  of about $-20^{\circ}$ declination; discontinue use of Schmidt plate data.

\item[Systematic error corrections:] a final tweak of the magnitude
  equation corrections of the CCD data brings the positional 
  system of UCAC4 closer to the UCAC2 system.

\item[High proper motion stars:] were identified in the UCAC 
   observational catalog and those positions used for UCAC4.

\item[APASS:] DR6 photometry in the B, V, g, r and i bands added.

\item[Photometric calibrations:] APASS data are used to calibrate 
    instrumental UCAC magnitudes over the entire dynamic range.

\item[Photometric bias:] as function of CCD $x$-coordinate removed.

\item[Bright stars:] added from FK6, Hipparcos and Tycho-2 catalogs.

\item[Hipparcos star numbers:] are linked to UCAC4 data which allows 
  easy inclusion of Hipparcos Catalogue data such as parallaxes.

\item[Cross reference:] to Tycho-2 and UCAC2 star catalog numbers added.

\end{description}

\section{CCD Data and Processing}

UCAC4 is based on the same pixel reduction as UCAC3 \citep{ucac3x}
and the same astrometric reduction pipeline from $x,y$ data to 
$\alpha, \delta$ of individual observations \citep{ucac3a} was 
used, except for bug fixes, a minor update of the magnitude 
equation correction, and adopting a lower threshold
to include more single observations of faint stars.

\subsection{Positions}

Positions in UCAC4 are on the ICRS
as realized by the Tycho-2 catalog, which was used
as reference star catalog in a conventional, frame-by-frame,
astrometric reduction after various corrections were applied,
utilizing the 2MASS data \citep{2mass} and east-west flip observations
as described in \citep{ucac3a}.

Instead of the bin-by-bin corrections for the poor charge-transfer 
efficiency (CTE) effect on positions as used for UCAC3, a smooth 
function with magnitude was adopted for the UCAC4 reductions using 
the same calibration data (based on 2MASS) as before.
CTE corrections for the 200 and 40 sec exposures (mainly used for the
extragalactic link program) as well as for 5 and 10 sec exposures 
were not applied correctly for UCAC3.  
This bug has been fixed for UCAC4, affecting only few CCD frames.
Overall the astrometric accuracy of UCAC4 observational positions
has not changed significantly from the UCAC3 release.

\subsection{Completeness}

The missing stars and multiple entries of some stars in UCAC3
were traced back to a bug in the merge stage of individual observations
to mean CCD-based positions.  This problem was corrected and a limit of 
2.0 arcsec was imposed to combine images to a single star.
This change lead to more stars with a blended image flag in UCAC4 than
before; however, it avoids listing unrealistic close, ``resolved" 
companions of unreal double stars.
After checking sample double stars discovered in UCAC data with the
26-inch speckle camera, potentially new real double stars were 
identified in UCAC4 and will be published separately.

For UCAC4, an even lower threshold was adopted for
faint observed images to enter intermediate star lists,
including objects with positions based on center-of-mass centroid
instead of a successful image profile fit.
However, these objects entered the UCAC4 catalog only if either a match 
with at least another observation existed, or a 2MASS match could be 
established, or a reasonable proper motion could be obtained by
matches with early epoch data.  
Whenever no formal position error from the CCD data of a star
could be derived (no successful image profile center fit,
for whatever reason) the positional error was set to 900 mas 
(1 pixel), which leads to early mean epochs and large proper 
motion errors when combined with other catalogs to arrive at 
the UCAC4 compiled catalog data.
This approach allowed the completeness 
level of the UCAC4 catalog to be pushed to a new high with an
estimated 110 million real stars (and about 92,000 galaxies),
but also allowed some artifacts to enter UCAC4,
estimated to be on the few percent level, particularly near
bright, overexposed stars.  

As with UCAC3, overexposed stars were propagated in the catalog 
production pipeline for reasons of completeness.  
For those stars, and other problematic 
images, the image center fit often failed, which is indicated in the
catalog by the number of ``used images'' being zero.  In these cases
no fit position could be obtained, instead the provided position is
only approximate, based on the centroid (first moments) of the light
distribution in the pixel data.
Contrary to UCAC3, all stars brighter than observed magnitude 8.5
in UCAC4 were compared to external catalogs and astrometric data
from these were substituted when deemed more reliable than UCAC
observations (see section 6.2).

\subsection{Photometry}

As with UCAC3, UCAC4 
gives 2 observed magnitude estimates, based on the volume of the
image profile model fitted, and a true aperture photometry.
Extinction coefficients are derived for each
exposure with respect to Tycho-2 stars adopting a linear model
with B$-$V color.  Thus a photometric zero-point was determined
for each CCD exposure and applied to the instrumental magnitudes
to arrive at our bandpass magnitudes based on the available
Tycho-2 stars in a given field.  

Early APASS DR2 data of 9 million stars distributed all over the sky
were used for photometric calibrations of UCAC-observed CCD magnitudes.
Contrary to previous releases, a photometric bias as a function of
the pixel $x$-coordinate was removed in the instrumental magnitudes
before constructing the UCAC4 catalog.
This bias is caused by the poor CTE performance of the detector
and affects the model and aperture magnitudes similarly, but not
identically.  The bias is also a function of magnitude and
slightly depends on exposure time.
Fig.~2 shows examples for 2 ranges in brightness.
A linear model was adopted to correct for this bias with respect to 
the center of the CCD field.  Different slope parameters as a function
of $x$-pixel coordinate were applied for NOFS and CTIO data, 
as well as for long and short exposures and binned by magnitude group.
Linear interpolation by magnitude was adopted for applying corrections.
The largest effect of this bias is about $\pm0.2$ mag.

The following procedure was performed separately for UCAC4 aperture
and model magnitudes to correct for the non-linearity of the UCAC
instrumental magnitudes.
A preliminary UCAC magnitude was calculated after applying the $x$-pixel
bias correction and using Tycho-2 stars for the photometric zero point.
This UCAC preliminary magnitude was then compared to the APASS r magnitude
and magnitude differences plotted as a function of APASS (V$-$r),
see Fig.~3.
This color-color diagram was fitted with linear and second order polynomials
in sections to allow calculation of an estimated UCAC-bandpass magnitude 
when only APASS V and r are given.
Very blue (V$-$r $\le$ $-$0.1) and very red (V$-$r $\ge$ 1.0) stars
were excluded in this process.
The differences between UCAC4 observed and UCAC-bandpass estimated
magnitudes (from APASS V,r) as a function of UCAC observed magnitude 
(Fig.~4) give the desired corrections to the preliminay UCAC magnitudes
to make them linear with respect to the APASS photometric system.
A zero-point offset is included in these corrections to force
the resulting UCAC$-$APASS(r) color to be zero for (V$-$r) = 0.

As noted in earlier releases, the largest deviations of the UCAC
magnitude scale is seen at bright magnitudes near saturation,
requiring a correction on the order of 0.5 mag.
The faint end needed a correction of about 0.3 mag.
For UCAC4, the use of the APASS data allowed 
these magnitude scale and offset corrections to be determined
much more accurately and reliably than in earlier releases.

Finally, the corrected UCAC magnitudes were run
through the Tycho-2 zero-point routine again to arrive at the
calibrated magnitudes of individual CCD frames.
Combining individual CCD frame photometry to mean catalog magnitudes
and calculating error estimates were performed in the same way as for 
UCAC3 \citep{ucac3r}.
Remaining sytematic errors in UCAC4 observed magnitudes are
estimated to be about 0.05 to 0.1 magnitudes, because 
many UCAC observations were performed under non-photometric sky 
conditions, and the imperfection of the applied corrections.
The later is most pronounced around magnitude 9 as for example
can be seen as slight discontinuity in the Fig.~1 histogram.
This is a magnitude calibration issue and not a missing star issue.

\section{Proper Motions}

The UCAC4 catalog heavily relies on the AC2000.2 Astrographic Catalogue
\citep{ac2000}, unpublished data from over 5000 astrograph plates measured 
with the StarScan machine \citep{starscan}, the NPM \citep{npm1}, \citep{npm2}, 
and the SPM \citep{spm2}, \citep{spm4} data.
All procedures to derive the UCAC4 proper motions remained the same 
as for earlier releases, see for example the UCAC1 paper \citep{ucac1}.
Most of the early epoch catalog used to derive UCAC4
proper motions also remain the same as for UCAC3.
However, Lick Observatory Northern Proper Motion (NPM) 1st epoch
data (Lick1 catalog) were used for UCAC4,
and field zero-point corrections for the Yale-San Juan Southern
Proper Motion (SPM) 1st epoch data (YSJ1 catalog) were applied. 
Both sets of plates were digitized
at the Precision Measure Machine (PMM) \citep{pmm} at NOFS, with
subsequent data reductions as a joined effort by Yale University and 
USNO, Washington \citep{ucac3r}.
Thus, for UCAC4, no Schmidt survey plate data were used at all,
which lead to a significant improvement in the UCAC proper
motions north of the SPM sky area ($\delta \ \ge \ -20^{\circ}$).

As with UCAC3, only the first epoch, blue SPM plate material was
utilized for UCAC4 proper motions of faint stars in the south.
In the meantime, the Yale astrometry group completed their reductions 
and derived a catalog of positions and absolute proper motions 
based on early and recent epoch SPM project observations, the SPM4,
\citep{spm4} utilizing galaxies to establish the proper motion
zero-point.
Due to the common first epoch data, even with differences in the
reduction algorithms, the SPM4 and UCAC4 proper motions will be 
correlated.
Table 2 provides statistics about matches of UCAC4 stars with
other catalogs as used to derive proper motions.

\subsection{NPM data}

All applicable Lick Observatory Northern Proper Motion (NPM)
first epoch plates (blue sensitive emulsion) were used to
construct a star catalog with mean epoch near 1950 covering
the about $-25^{\circ} \le \delta \le +90^{\circ}$ area of sky.
These were among the earliest plates scanned on PMM and no
pixel data could be saved at the time.  The provided data 
were the $x,y$ pixel locations of stellar images measured 
on individual PMM CCD ``footprints" with the PMM pipeline
mapping parameters already applied.

In order to be able to run these data through the StarScan
\citep{starscan} pipeline, the previously applied PMM modelling 
was removed to obtain raw $x,y$ pixel centroid data.  
The StarScan pipeline then generated new, global $x,y$
data on the coordinate system of a plate using improved modelling.

The global $x,y$ data then was sent to the Yale astrometry group
for further astrometric reductions to obtain $\alpha, \delta$
coordinates on the sky, applying elaborate systematic error
corrections by utilizing the various orders of grating images
and 2 sets of exposures in each field.
These procedures follow closely the reduction process of the
SPM plates \citep{spm4}, with some modifications due to differences
in magnitude overlaps of grating images between SPM and NPM data.
The resulting, unpublished catalog ``Lick1" of over 168 million stars 
and galaxies was sent to USNO, Washington, to provide the early epoch
positions of faint stars for UCAC4 proper motions. 

\subsection{SPM and NPM corrections}

Despite the great care taken to control systematic errors as a
function of magnitude and $x,y$ location, residual magnitude 
equations remain in the NPM and SPM first epoch data, the Lick1
and YSJ1 catalogs.
First, the field-dependent systematic errors were corrected,
then the overall zero-point of the proper motions per NPM and SPM
field.
All UCAC stars with preliminary proper motions were associated with 
the nearest field centers of NPM and SPM plates.
A tangential plane projection then provides the $x,y$ location
of UCAC stars on NPM and SPM plates.

\subsubsection{Field dependend corrections}

The preliminary UCAC proper motions, derived only from the CCD
data mean positions and NPM or SPM first epoch positions were
binned (0.5 degree) on the NPM and SPM plate pattern separately 
for various declination zones but combining data from several fields
along right ascension (RA).
Mean proper motions per $x,y$ bin were calculated excluding
the high and low 20\% of proper motions in each bin.
Fig.~5 shows an example of such a vector plot of artificial 
mean proper motions in the tangential plane of NPM fields.
Excluding 10\% or 30\% (instead of 20\%) of the high and low 
proper motions in this process yields almost identical results.
Typical systematic error vectors are 0.5 to 2 mas/yr with
a formal error of about 0.12 mas/yr on average for the NPM
fields and somewhat smaller corrections for the SPM data.

No physical reason exists for such a pattern in the average
proper motions of stars on the sky on such small scales ($\approx$ 
1 degree), repeating with the pattern of NPM and SPM fields.
Thus, the differences in proper motions seen here between bins
are systematic errors in the NPM and SPM data.
Differences in proper motions from bin to bin were translated 
by the known epoch difference to positional offsets at the respective
NPM or SPM epoch and corrections applied to the first epoch
NPN and SPM catalog positions respectively, keeping the proper
motions at the field centers unchanged.
Separate patterns were derived for these corrections per 
declination zone.  Variations as a function of declination were
found to be insignificant and a combined pattern was established
for NPM and SPM respectively, which was used to apply the
corrections.
These corrections are dominated by faint stars in the about
14 to 16 mag range with an average absolute correction of about
1 and 0.25 micrometer for the NPM and SPM fields, respectively,
with the largest corrections about 3 times larger than the average.

\subsubsection{Proper motion zero-point corrections}

Here, we follow a procedure adopted for the construction of
the SPM4 catalog \citep{spm4}.
Besides systematic errors as a function of $x,y$ location on 
NPM or SPM plates, an overall magnitude equation (field by field)
was found when looking at proper motions of galaxies.
In order to utilize as many galaxies as possible, the entire Lick1 (NPM)
and YSJ1 (SPM) catalogs were matched with the 2MASS extended source
catalog, not just those objects present in UCAC.
Position differences between 2MASS and Lick1 catalog positions were 
calculated for over 678,000 galaxies with formal position errors
less than 400 mas and averaged by Lick1 field.
The positional offsets of typically 100 mas are attributed to systematic
errors in the Lick1 data at an average magnitude around R = 16.

Similarly, average position differences between Lick1 and the Hipparcos
Catalogue (using Hipparcos proper motions) were calculated field-by-field
for a total of over 68,000 acceptable stars at the Lick1 epoch.  
Again, the position offset was attributed to systematic errors in the 
Lick1 data at a mean magnitude of about 7.5.
Of the 1390 NPM fields in Lick1, a total of 1347 were found with
sufficient data from both the galaxies and Hipparcos stars.
For the remaining fields, no corrections were applied.
An example of the results is shown in Fig.~6 for a range of fields
and the $x$-coordinate (along RA).
The position differences with Hipparcos are less pronounced than
in this example for the $y$-coordinate (declination).
Results for the SPM (YSJ1) data are similar with pronounced
systematic offsets for the Hipparcos data ($\pm$ 200 mas) and
low significant, small offsets for the galaxy data, similar to NPM.

A linear magnitude equation in the Lick1 data was assumed between
the position offsets at the bright end (Hipparcos stars) and faint
end (galaxies).  The full correction, as derived from the above
procedure, was applied whenever the formal error on the position
offsets (for Hipparcos stars and galaxies separately) is less than 
50 mas.  For larger errors per field and coordinate the above derived 
corrections were scaled down by a factor of 50 mas divided by formal 
error.  Then, the position difference between bright and faint offsets
was divided by the 8.5 magnitude difference to arrive at the
magnitude equation slope.  Positions of all Lick1 objects were then 
corrected differentially for this magnitude equation to force the 
position offsets (Lick1 - Hipparcos, Lick1 - galaxies) to zero.
The YSJ1 (SPM) data was handled in the same way.

\subsection{High proper motion stars}

Stars with high proper motions (HPM) were handled specifically.  
In the north the LSPM-North Catalog \citep{LS2005}  
of 61977 new and previously known high proper motion stars having
proper motions greater than 0.15"/yr was used. In the south, many smaller
surveys along with the Revised NLTT Catalog \citep{SG2003} were
used, which produced 17730 unique high proper motion stars greater than
0.15"/yr. In both the north and south a supplemental list of proper
motion stars greater than ~0.15"/yr from the Tycho-2 and Hipparcos
catalogs was used to fill in any gaps. In chronological order, the
smaller southern surveys used include: (1) 7 papers covering various
portions of the southern sky by Wroblewski and collaborators
(Wroblewski \& Torres 1989, 1991, 1994, 1996, 1997; Wroblewski \& Costa
1999, 2001), (2) UK Schmidt Telescope survey plates of 40 survey
fields by Scholz and collaborators \citep{Scholz2000}, (3) The
Calan-ESO survey \citep{Ruiz2001} (4) SuperCOSMOS-RECONS
proper-motion survey of the entire southern sky (Henry et al. 2004;
Subasavage et al. 2005a, 2005b; Finch et al. 2007; Boyd et al. 2011),
(5) the Southern Infrared Proper-Motion Survey (SIPS; Deacon et
al. 2005), (6) Lepine's SUPERBLINK survey of a portion of the southern
sky \citep{Lep2008} and (7) UCAC3 proper motion survey 
(Finch et al. 2010, 2012).

Then, we identified these stars in our CCD observations using a 2-step
approach.  For each individual exposure we established a list of HPM stars
which could be present in that field.  HPM star positions were calculated
for the epoch of that exposure and then matched with the individual RA,Dec
observations of that exposure to identify and flag HPM stars on each
exposure (object type = 3).

Contrary to UCAC3, a UCAC4-based solution for mean position
and proper motion was attempted for all stars, including the HPM stars.
The position and proper motion solution obtained by the above procedure
was substituted by zero proper motion and the mean CCD data position at
mean observational epoch for the following cases:
\begin{itemize}
\item proper motion (PM) solution failed or showed large errors
   ($\ge$ 500 mas, $\ge$ 50 mas/yr)
\item derived PM is larger than 500 mas/yr in either component
\item derived mean epoch is earlier than 1947
\item difference between mean catalog position and mean CCD data 
   position is $\ge$ 3 arcsec
\end{itemize}

The mean CCD position of a star is the UCAC observed position obtained 
by combining data of individual exposures of that star.
Stars with early mean epoch are problematic; including poor CCD
position data or possible mismatches across involved catalogs.
The stars cut by the above criteria are thus added to the group of 
``no proper motion" stars, i.e.~those that did not match up with other 
catalogs to even begin the proper motion calculation.  
All stars were then checked against the external set of HPM stars.  
The PM from the external catalog was used for stars with no UCAC4 
PM solution and for those where the difference in PM for either 
component exceeded 40 mas/yr.  Thus, we trust the external catalog 
data more than the UCAC4 derived proper motions in those cases.

\section{Comparisons with other Catalogs}

For the following comparisons with the UCAC4 release data,
only stars with unique, single matches to the respective
catalogs were used.  A match radius of 2.0 arcsec
was adopted for positions at the desired common match epoch,
by applying proper motions as specified below.
Tests with a match radius of 1.0 arcsec gave almost identical results.

\subsection{UCAC2}

Figures 7 to 12 illustrate the systematic differences between the
UCAC4 and UCAC2 data releases regarding magnitude and positions,
for the southern and northern hemisphere separately.
Almost all of the 48.3 million entries in UCAC2 were matched with UCAC4.
For the figures we excluded flagged double stars as well
as stars with a formal position error larger than 150 mas.

The differences in photometry between UCAC4 and UCAC2 display 
small scatter but complex, large systematic offsets (Figs.~7, 8).
The UCAC4 photometric system is expected to be significantly
better than the UCAC2 data due to the better calibrations.
Figs.~7 and 8 show results for aperture photometry, while results
for model magnitudes are very similar.
Largest differences are seen at the bright end where UCAC2 data
suffers from uncalibrated non-linearities near saturation,
which already were removed in UCAC3.
Small discontinuities on the 0.02 mag level seen in Fig.~8 are
not of any concern because UCAC magnitudes will have larger
local systematic errors anyway.

Figs.~9 and 10 show the systematic position differences (at epoch 2000)
between UCAC4 and UCAC2 for the southern and northern hemisphere,
respectively, as a function of magnitude.
Systematic differences are only a few up to 10 mas over the
entire range.
The relative differences (shape of these patterns) is determined 
mainly by differences in the CTE calibration models used for 
the data sets, while the absolute zero-point is determined by the 
mean of Tycho-2 stars around magnitude 10 to 11 (common system
between UCAC2, UCAC4, and Tycho-2 by design).

Fig.~11 and 12 show the UCAC4$-$UCAC2 position differences as a
function of declination.  The mean offset in these figures is 
determined by the mean offset in the previous figures at faint
magnitudes, where the majority of stars are.
Local variations in these data are very small (few mas) with
the exception of $\Delta\delta$ around +50$^{\circ}$, where
the UCAC2 data runs out and only a small range in RA is being
averaged over.

\subsection{2MASS}

Figures 13 and 14 show position differences between UCAC4 and 2MASS
as a function of magnitude, for the southern and northern hemisphere,
respectively.
The UCAC4 proper motions are used to bring the UCAC4 positions to
the 2MASS epoch (about 1998 to 2002) for each individual star matched 
uniquely within 2 arcsec.
Figures 15 and 16 show these position differences as a function of
declination.
Each dot represents the mean over 3000 stars, excluding stars with
a UCAC4 double star flag and those with an estimated position error
larger than 150 mas.  This exclusion effectively cuts stars fainter 
than about 16.5 mag.
The systematic UCAC4 minus 2MASS position differences are typically
about 10 mas, with additional local systematic variations.
The sawtooth pattern can be explained by residual
systematic errors in the UCAC4 proper motions, based on Lick1 and YSJ1
data (5 degrees is the size of individual NPM, SPM plate fields).
For those first epoch data no position averaging over plate boundaries
was performed.
Plots similar to Fig.~13 and 14 comparing 2MASS with UCAC3 and UCAC2 
were presented in the UCAC3 release paper \cite{ucac3r}.
We note that the position system of UCAC4 agrees better with 2MASS 
than did UCAC3. This is the result of a better magnitude dependent
correction (CTE) for faint stars in UCAC4 versus UCAC3.
Note, although 2MASS positions were not directly used in the UCAC4
final astrometric ``plate" solution, 2MASS data were used to correct
some magnitude dependent systematic errors in UCAC4 data.
Thus 2MASS and UCAC4 are somewhat correlated.

\subsection{Proper motion catalog comparisons}

To highlight possible systematic errors in proper motions between 
various catalogs, a random slice of the sky between right ascension 
6.0 and 6.1 hours was picked.  
Such a slice is narrow enough to sample parts of
individual first epoch plates instead of averaging over an entire zone
or several plates.  Thus, limitations in astrometric calibrations
as a function of the location of stars on the $x,y$ plane of a plate 
will become apparent.

The UCAC4 was matched with the PPMXL \citep{ppmxl},
XPM \citep{xpm}, and SPM4 \citep{spm4} catalogs and differences
in proper motions are plotted as function of declination in
Figs.~17 to 19.  
The range of $-60^{\circ} \ \le \ \delta \ \le \ -30^{\circ}$ is
choosen to give sufficient coverage from the SPM data.
The difference between XPM and PPMXL can be inferred by taking
the differences between the UCAC4$-$XPM and UCAC4$-$PPMXL plots.

All catalogs are somewhat correlated, for example, the UCAC4 and SPM4 
share some first epoch astrograph data, and XPM and PPMXL share 
Schmidt plates survey data.  Nevertheless, significant 
systematic differences in proper motions between all catalogs 
typically on the 2 mas/yr level, up to about 6 mas/yr locally
are apparent, with the UCAC4 vs.~SPM4 differences being somewhat 
smaller.

\subsection{Other external checks}

UCAC4 proper motions in RA vs.~proper motions in Dec is plotted
in Fig.~20 for a field around the open cluster M67.  Stars within 
20 arcmin of the cluster center and with formal, random errors in 
proper motion $\le$ 7 mas/yr per component (which excludes 4\% of 
the stars) are plotted.  
From these UCAC4 data, the mean proper motion of the cluster is 
found to be about $\mu_{\alpha \cos \delta}$ = $-$9.5 mas/yr
and $\mu_{\delta}$ = $-$4.5 mas/yr with an estimated formal error of
about 1 mas/yr (and expected systematic errors somewhat larger than
that).  These values compare very well with the published absolute 
proper motion for M67 ($-$9.6 and $-$3.7 mas/yr) from a recent paper
\citep{Bel2010} based on CFHT data and tied to galaxies.

\section{The Catalog}

\subsection{Main zone files}

The UCAC4 data files are organized in 0.2 degree wide declination
zones, numbered from 1 to 900 beginning at the South Celestial Pole. 
Within each zone, stars are sorted by ascending right ascension.
These 900 zone files are binary, containing 78-bytes fixed length
records with integers for each catalog entry (mostly stars).
The byte order is that of the native Intel-type processor binary
data format.  For some computers, a byte-swap might be needed.
Table 3 describes all data items for each star.
Detailed remarks are given in the readme file, which comes
with every data distribution (DVD or online as Vizier catalog 
I/322 at CDS).
Sample access code (in Fortran and C) is provided as well as 
index and other auxiliary files.
The UCAC4 distribution on DVD furthermore contains all-sky
plots showing catalog density and mean proper motions, historic 
information such as relevant papers, presentations and a
snapshot of the UCAC Web pages including pictures.
The online version provides real numbers in ASCII format versus the 
original integer data and gives declination in degree while the
DVD release gives south pole distance in mas (Table 3).

The official star identification name is of the format
UCAC4-zzz-nnnnnn.
The 3 digit ``zzz'' number is the zone number for a star,
while ``nnnnnn'' is the 6 digit record number for that star
along its zone file (with leading zeros, if needed).

Similar to previous releases, UCAC4 is a compiled catalog giving the
weighted mean position and proper motion of stars based on all input
catalogs, including the CCD data.
Subject to available resources, USNO plans to release the individual
CCD observations (positions at epoch).  To obtain this about 50 GB
large dataset, please contact the lead author.

UCAC4 does contain about 92,000 galaxies.
No star/galaxy separation parameter based on pixel data is present
in UCAC4.  However, flags are provided that indicate matches
with known non-stellar objects, from the 2MASS extended source 
catalog, the LEDA galaxy catalog, and non-stellar flags copied 
from the SPM data.

\subsection{Supplement stars and data}

Contrary to previous releases, UCAC4 was supplemented by bright
stars in the attempt to provide a catalog complete from the
brightest naked eye stars to about R = 16.
First, information from the Tycho-2 and Hipparcos (including annexes)
catalogs were merged with FK6 \citep{fk6} data.
In order of priority,
astrometric data are taken from FK6 if available, else from the
2007 Hipparcos release \citep{hip2}, else from the original 
Hipparcos Catalogue \citep{hipcat}, else from Tycho-2 \citep{tycho2}.
This combined catalog of over 2.5 million stars was matched with
UCAC4 and stars not found in UCAC4 observational data within 2 arcsec
were added to the UCAC4 release catalog.
Also, a total of 2 high proper motion stars was added manually.
For these supplemented stars as well as for all UCAC4 entries identified
as Hipparcos stars, a unique star ID number (column 51 of the zone files)
of below 1 million was assigned and additional data provided in a
separate table indexed with that star number.
Thus most of the Hipparcos Catalogue data (including parallaxes)
are linked to UCAC4 entries.

A cross reference to the Tycho-2 catalog was created and added to
the release in a separate file.  This file contains 2,549,788
entries linking the original Tycho-2 star number (3 parts) to the
official UCAC4 star number (zone number and record along zone) and
the main data table column 51 entry (single, unique star ID number).

\section{Discussions and Conclusions}

The UCAC4 is the final release of this project, providing an
all-sky, astrometric catalog to R=16 magnitude.
Fig.~21 shows the distribution of mean epochs of UCAC4 stars.
The recent epoch CCD observations dominate, while for some stars
the mean epoch can go back several decades.
For these stars the weight of the CCD observations used in the
calculation of proper motion is low, which can happen due to a
number of reasons like faintness of star or elongated image due to
multiplicity. 

Fig.22 shows the distribution of UCAC4 proper motions
(here for the RA component) on a logarithmic scale.
The declination proper motion distribution looks very similar. 
The overdensities near 100 and 200 mas/yr are likely caused by 
contaminations of stars with mis-matched positions between
early and late epoch data at the cut-off match threshold.

Fig.~23 shows the distribution of UCAC4 formal position errors at
mean epoch, which peaks near 18 mas.  Most stars have formal
position errors of between 15 and 100 mas, depending on magnitude.
Real position errors at current epoch are larger due to propagation 
of proper motion errors and additional systematic errors.
The positions at mean epoch in UCAC4 are closer to the UCAC2
and 2MASS systems than those of the UCAC3 release.
UCAC4 also contains more stars and includes bug fixes that makes
UCAC3 obsolete now.

Fig.~24 shows the distribution of UCAC4 formal proper motion errors
which peaks at 4 mas/yr. 
The small overdensities near 18 mas/yr and 32 mas/yr are caused by
the adopted position error of 900 mas for no-fit CCD positions in
the proper motion calculation process.  The mean epochs of CCD
observations, SPM, and NPM data are about 2000, 1972 and 1950,
respectively which lead to about 32 and 18 mas/yr formal proper
motion errors.  Entries in UCAC4 with a formal proper motion error 
larger than about 12 mas/yr should be considered problematic and might
not correspond to real stars.

Systematic errors in proper motions are estimated to be on the few 
mas/yr level as indicated by external comparisons with other high
precision (nearly) global catalogs which all claim to be on the
same inertial system, the ICRS (Fig.~17 to 19).
 
2MASS near-IR photometry has been added to UCAC4 as with earlier
releases.  For the first time a large fraction (about half) of the
stars in UCAC4 now also list precise optical photometry in up
to 5 bands from the APASS DR6 including some single observations. 
APASS V and r magnitudes of DR2
were used to calibrate the UCAC4 observed magnitudes (Fig.~25),
which should be more reliable than in any previous UCAC release.
The UCAC4 magnitudes are near the Sloan r magnitudes and between 
V and r.

An effort was made to utilize block adjustment (BA) techniques \citep{BAnz}
for the final UCAC release.  However, simulations indicated that the
small size of individual CCD images combined with the only 2-fold
overlap pattern leads to very slow convergence.  In light of still
remaining systematic position errors in UCAC data on the 10 to 20 mas
level, much more elaborate simultions are needed to prove any
improvement of a BA solution of UCAC data over the classical solution
adopted for UCAC4.  Therefore the BA solution of the data has not 
been pursued further at this time.

Reductions of the extragalactic link data of UCAC is still in progress.
There are indications for both local reference star zonal errors
as well as blended images or host galaxy contamination of some sources
when looking at the optical$-$radio position differences of ICRF sources
(Zacharias, Zacharias \& Finch, 2012).
However, no revised UCAC release after UCAC4 is planned.
Significant improvements in global astrometry will come from new
observations.  The USNO Robotic Astrometric Telescope (URAT)
project is already underway 
(www.usno.navy.mil/usno/astrometry/optical-IR-prod/urat).



\acknowledgments

We would like to thank everyone involved in the monumental NPM and SPM
projects, in particular Bill van Altena, Arnold Klemola, Burton Jones
and Bob Hanson.
For a complete acknowledgement regarding the UCAC project, please
see the author list and acknowledgement sections in the UCAC2 and UCAC3
release papers.  
The following acknowledges only additional contributions
directly related to this UCAC4 release paper.

Sean Urban is thanked for his Hipparcos and Tycho-2 merged catalog,
which was augmented with FK6 data here and used to supplement bright
stars in UCAC4.
We are grateful for the many, constructive comments we received from
the following testers of the UCAC4-beta data, in particular
Rae Stiening (from the 2MASS project) and Dave Herald (Australia)
as well as Dave Gault (Australia), Rama Teixeira (Sao Paulo, Brazil),
Christine Ducourant (Bordeaux, France), Bill Gray (Project Pluto),
Steve Preston (minor planet occultation community), Ricky Smart 
(Torino, Italy), Sean Urban, Greg Hennessy, Paul Barrett, and
Bob Zavala (USNO).

Francois Ochsenbein from CDS, Strasbourg is thanked for preparing
the UCAC4 on-line release and the staff of CDS is thanked for
hosting all UCAC releases over the years.
Ralph Gaume is thanked for supporting the UCAC project as Head of
the Astrometry Department over the many years.
National Optical Astronomy Observatories (NOAO) are acknowledged for
IRAF, Smithsonian Astrophysical Observatory for DS9 image display software,
and the California Institute of Technology for the {\em pgplot} software.
More information about the UCAC and follow-up projects is available at 
\url{www.usno.navy.mil/usno/astrometry/}.
Finally, the referee (S.~R{\"o}ser) is thanked for valuable comments
which improved this manuscript.





\clearpage

\begin{figure}
\epsscale{1.00}
\includegraphics[angle=-90,scale=.34]{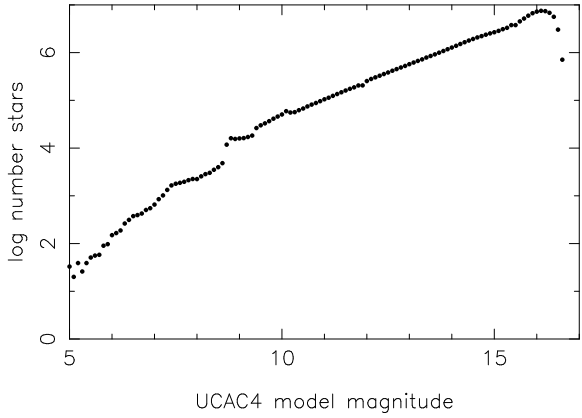}
\caption{Histogram of UCAC4 model magnitude distribution.
         The slight discontinuity near magnitude 9 is likely 
         caused by the saturation limit of the long exposure.}
\end{figure}


\begin{figure}
\epsscale{1.00}
\includegraphics[angle=0,scale=.42]{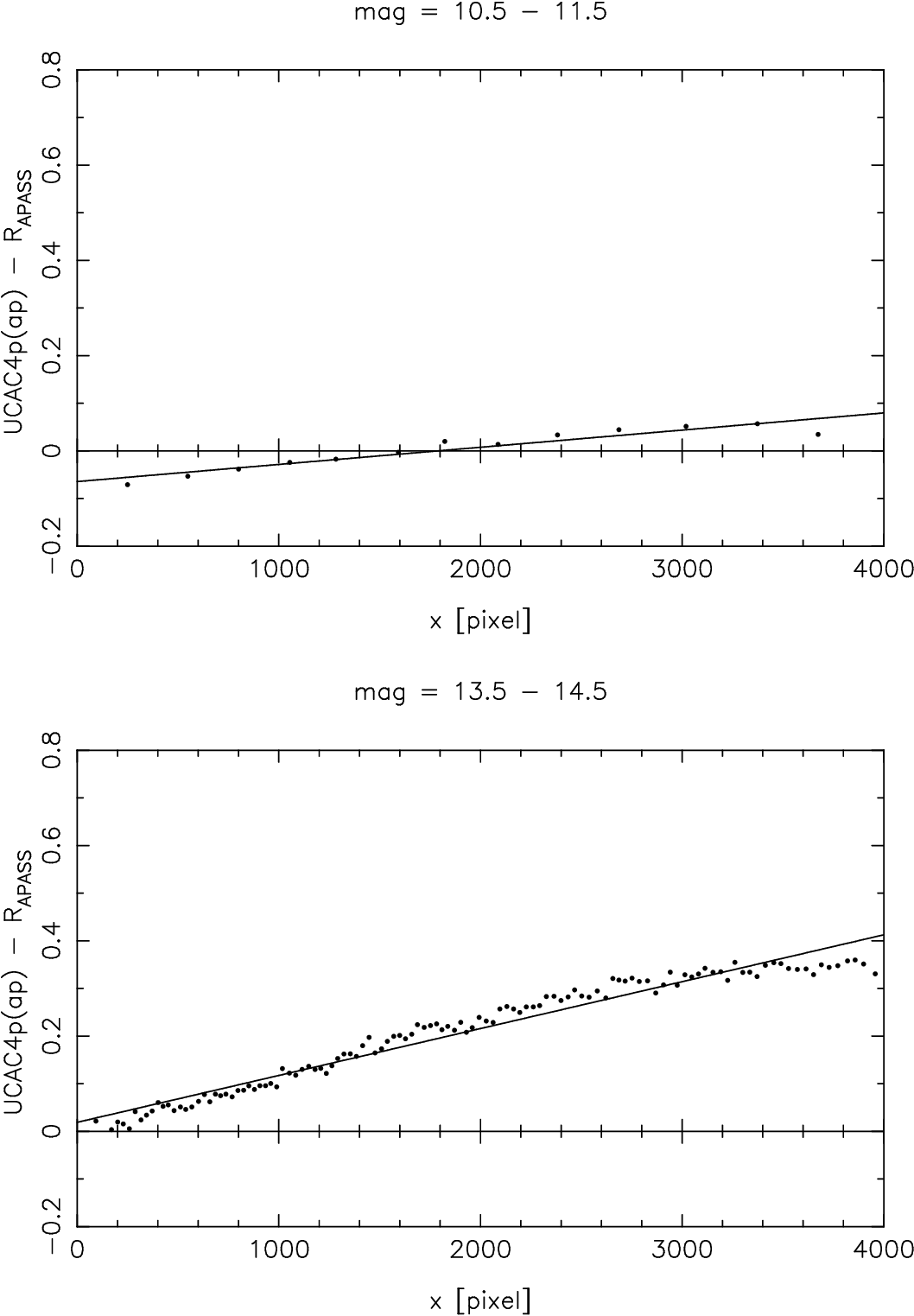}
\caption{Preliminary UCAC4 aperture magnitude minus APASS
         r magnitude as a function of $x$-pixel coordinate.
         Results are shown for 2 different brightness intervals
         with the adopted linear fit model superimposed.}
\end{figure}

\begin{figure}
\epsscale{1.00}
\includegraphics[angle=-90,scale=0.34]{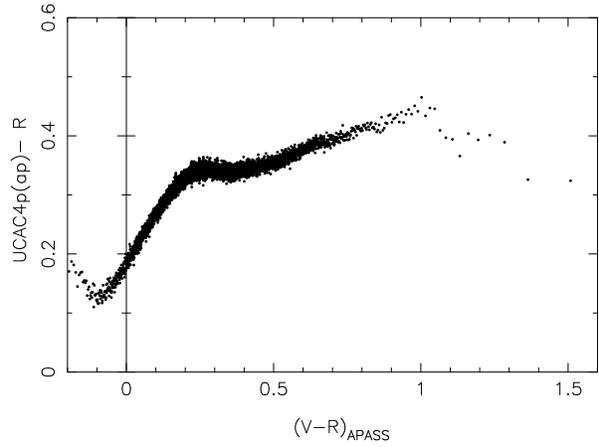}
\caption{Preliminary UCAC4 aperture magnitude minus APASS
         r magnitude as a function of APASS (V$-$r) color.
         This relationship allows an estimate of UCAC-bandpass
         magnitude from APASS V and r data only.}
\end{figure}

\begin{figure}
\epsscale{1.00}
\includegraphics[angle=0,scale=0.42]{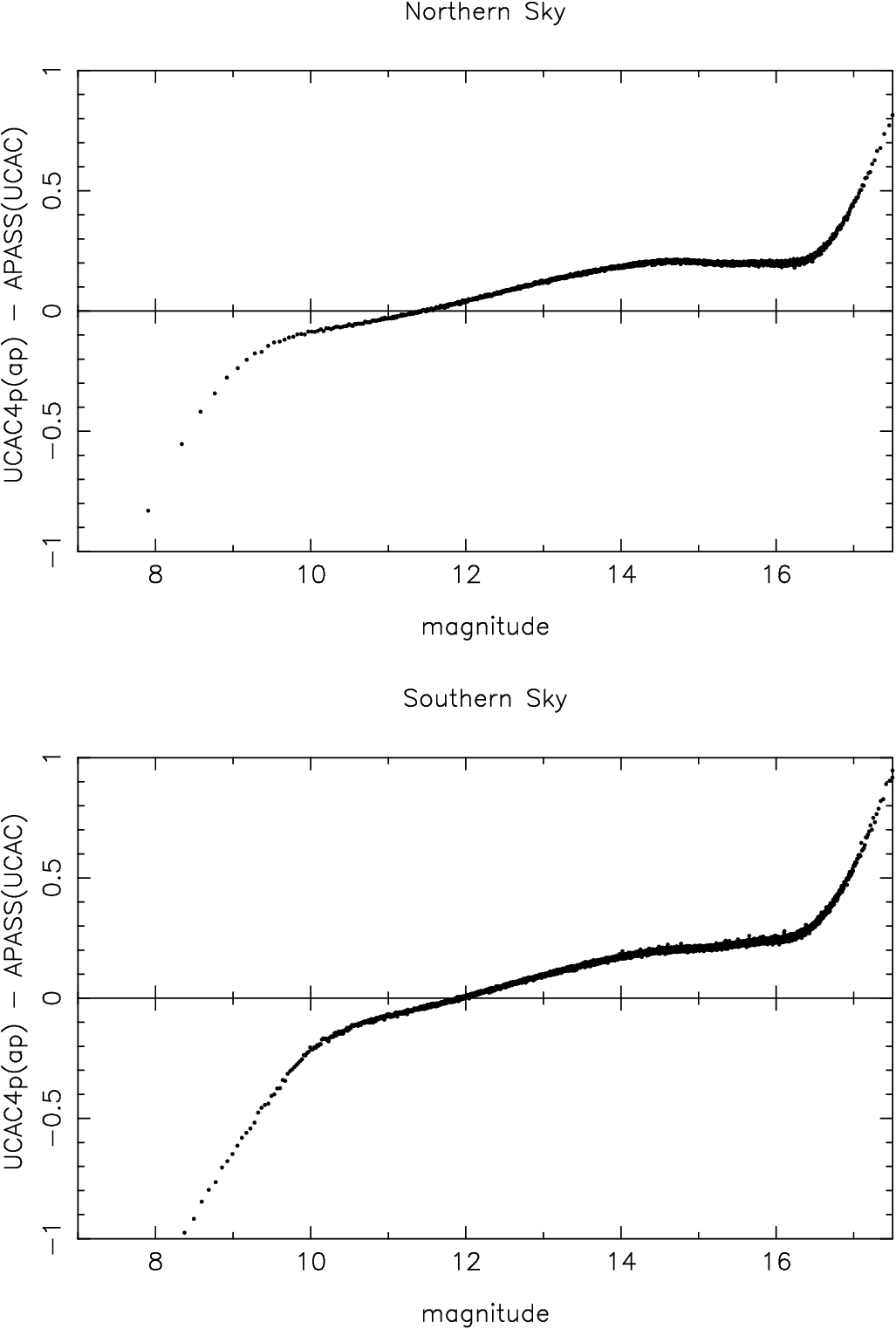}
\caption{Preliminary UCAC4 aperture magnitude minus estimated
         UCAC-bandpass magnitude (from APASS V,r data)
         as a function of preliminary UCAC4 aperture magnitude.
         These corrections are to be applied to the preliminary
         UCAC4 magnitudes to arrive at calibrated UCAC4 magnitudes.}
\end{figure}

\begin{figure}
\epsscale{1.00}
\includegraphics[angle=0,scale=.45]{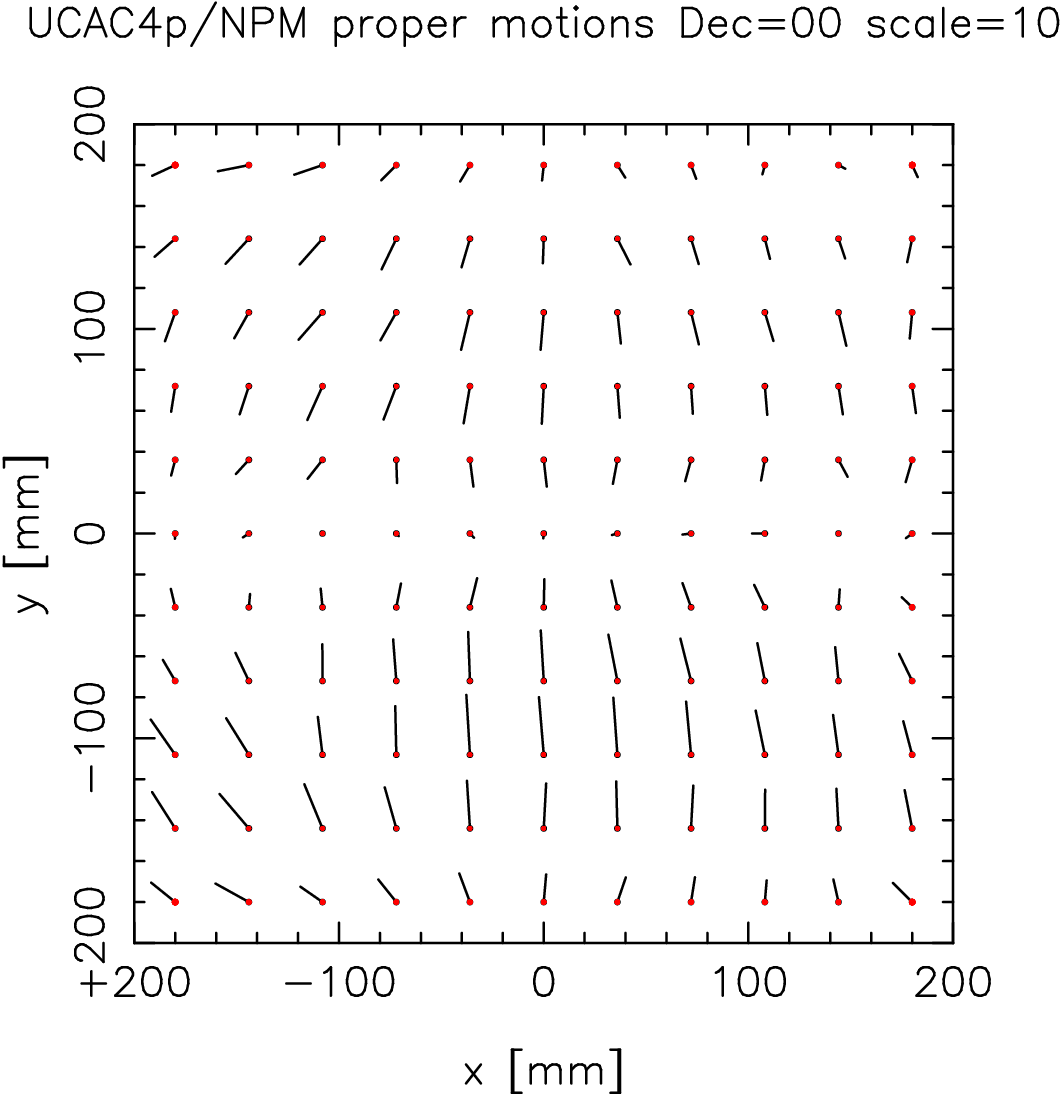}
\caption{Average preliminary ``proper motions" between UCAC epoch
         CCD observations and NPM first epoch data before
         corrections as functions of bins on the NPM plate
         tangential plane.  Data are averaged over 72 fields
         along declination = 0 degree, excluding 20\% of the
         high and low proper motions in each bin.
         The longest vectors are about 3 mas/yr.}
\end{figure}

\begin{figure}
\epsscale{1.00}
\includegraphics[angle=0,scale=.42]{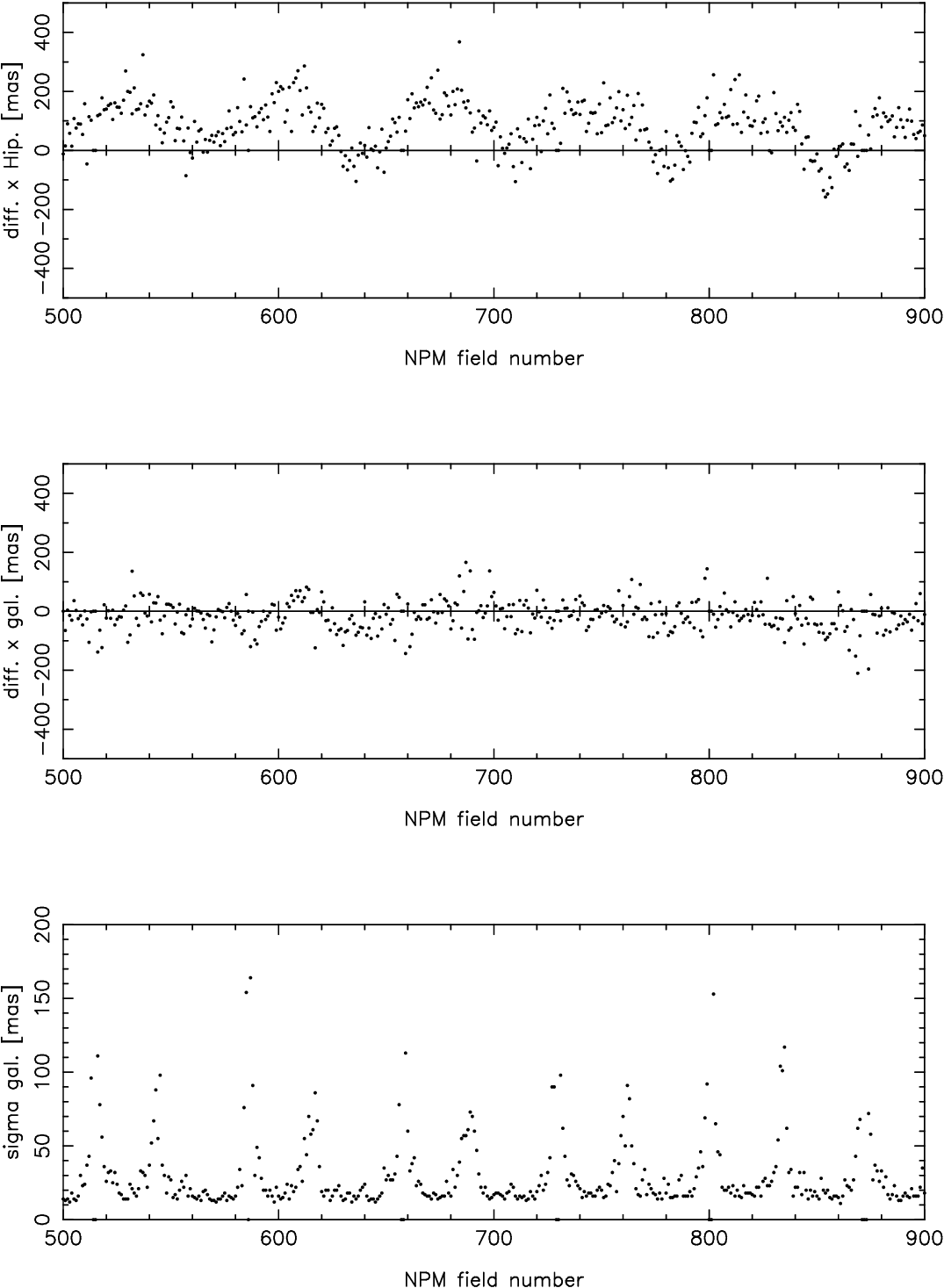}
\caption{Example of field-by-field mean position differences
         ($\Delta x = \Delta\alpha \cos \delta$)
         between NPM (Lick1) catalog (before corrections) and
         Hipparcos Catalogue positions at Lck1 epoch (top).
         The middle graph shows the position differences of
         Lick1 and 2MASS for extended sources (galaxies),
         and the graph at the bottom gives the formal error
         of the galaxy positions.  The formal error for the
         Hipparcos positions is about constant at $\approx$ 40 mas.}
\end{figure}

\begin{figure}
\epsscale{1.00}
\includegraphics[angle=-90,scale=.33]{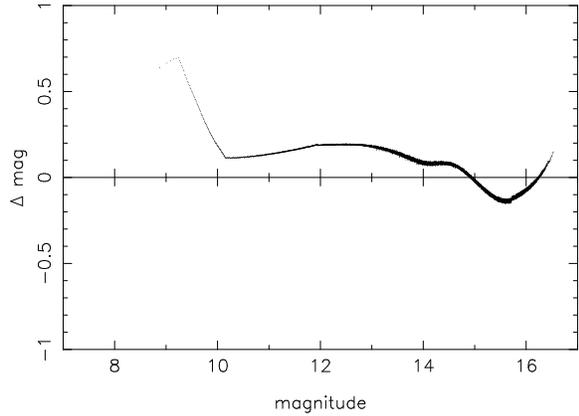}
\caption{Average UCAC4$-$UCAC2 differences in aperture photometry
         for southern hemisphere as a function of UCAC4 aperture magnitude.}
\end{figure}

\begin{figure}
\epsscale{1.00}
\includegraphics[angle=-90,scale=.33]{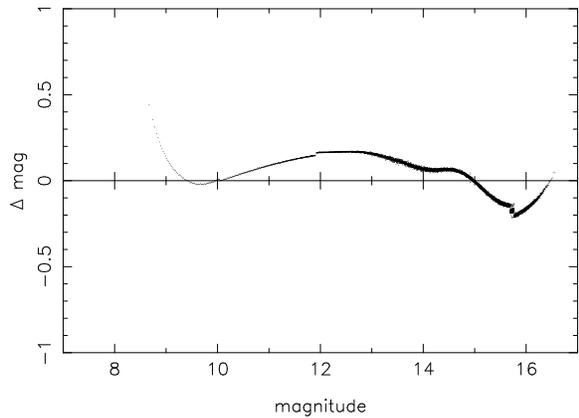}
\caption{The same as the previous figure for northern hemisphere data.}
\end{figure}

\begin{figure}
\epsscale{1.00}
\includegraphics[angle=0,scale=.42]{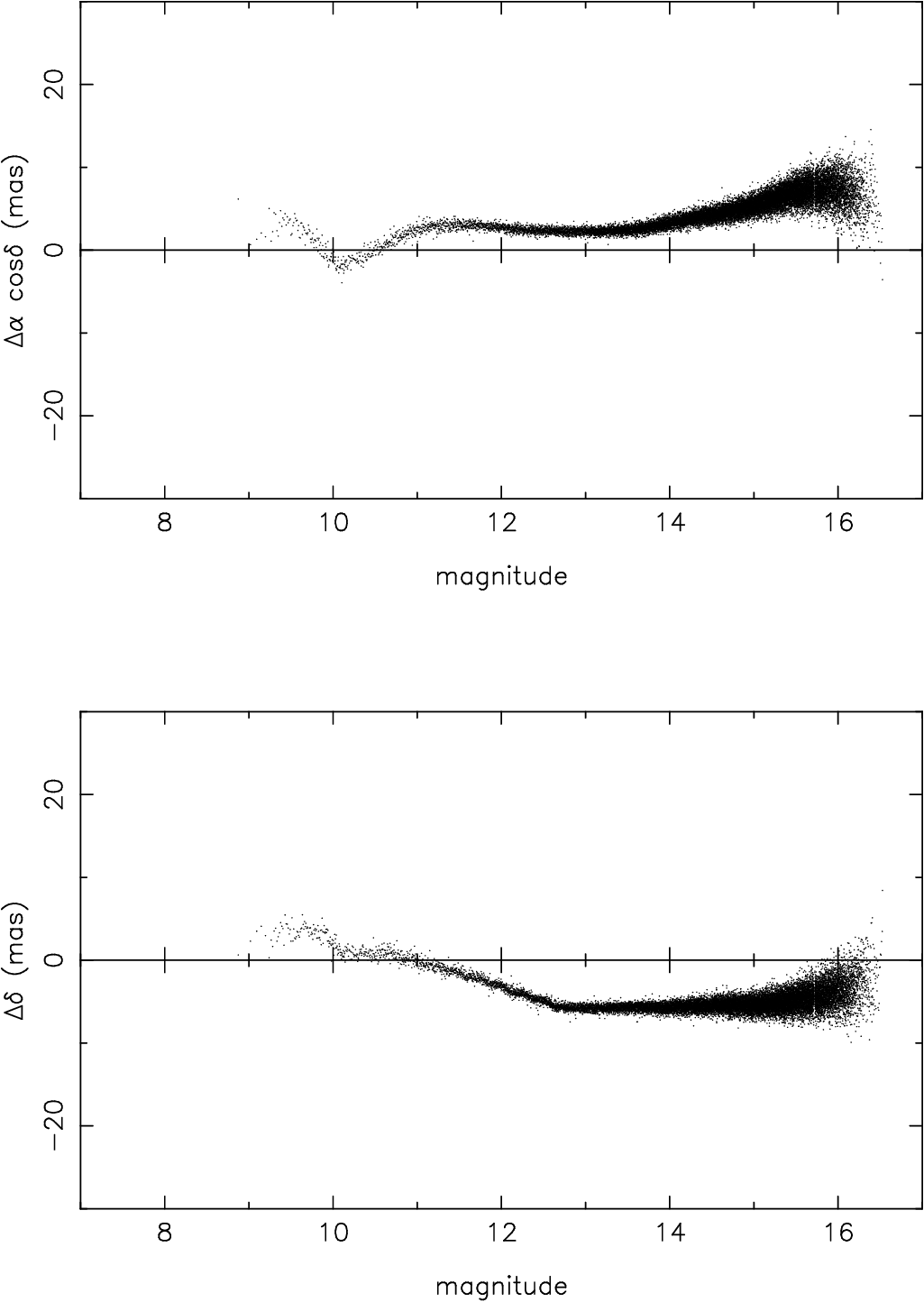}
\caption{Average position differences UCAC4$-$UCAC2 for
         southern hemisphere as a function of UCAC4 aperture magnitude.}
\end{figure}

\begin{figure}
\epsscale{1.00}
\includegraphics[angle=0,scale=.42]{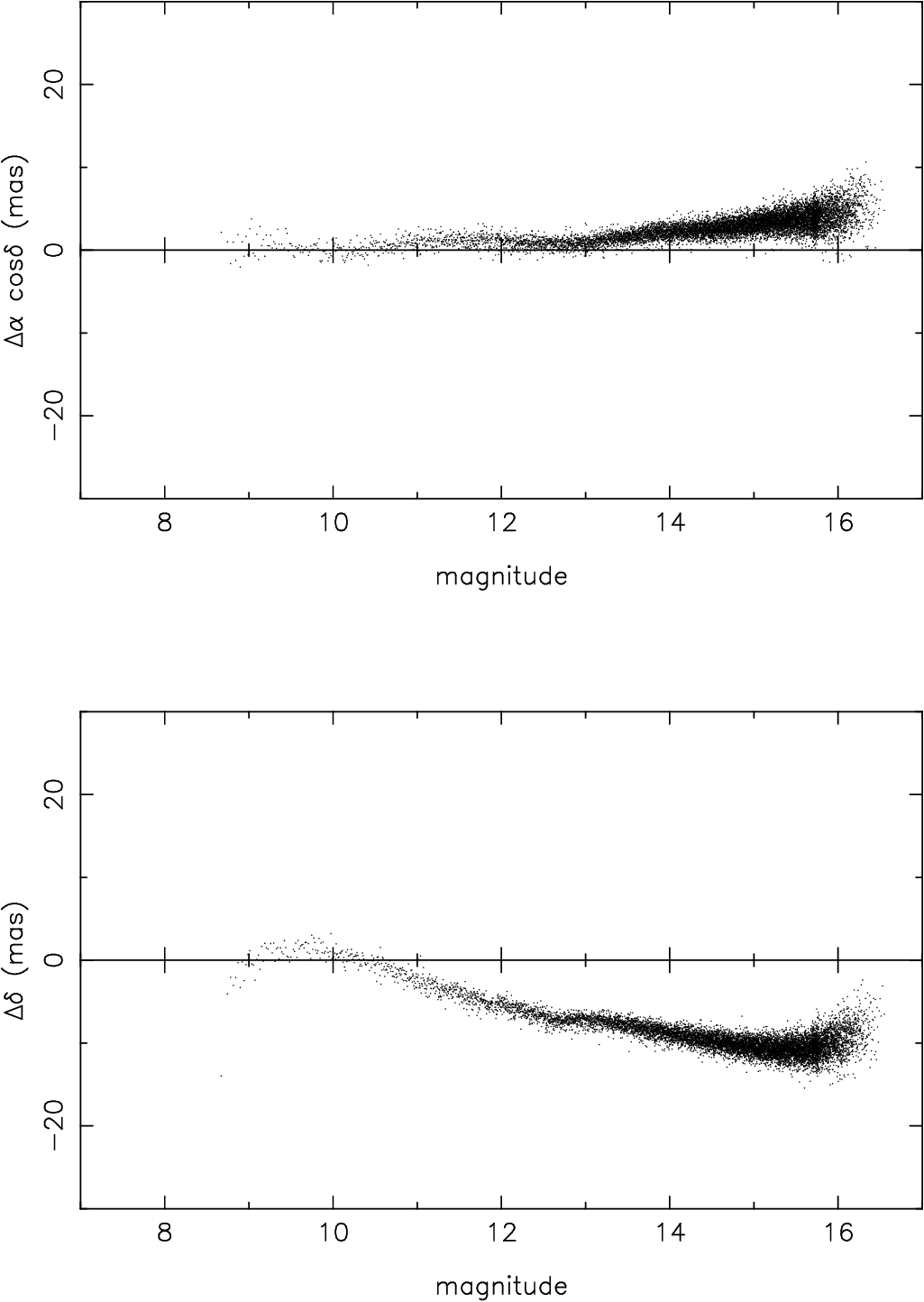}
\caption{The same as the previous figure for northern hemisphere data.}
\end{figure}

\begin{figure}
\epsscale{1.00}
\includegraphics[angle=0,scale=.42]{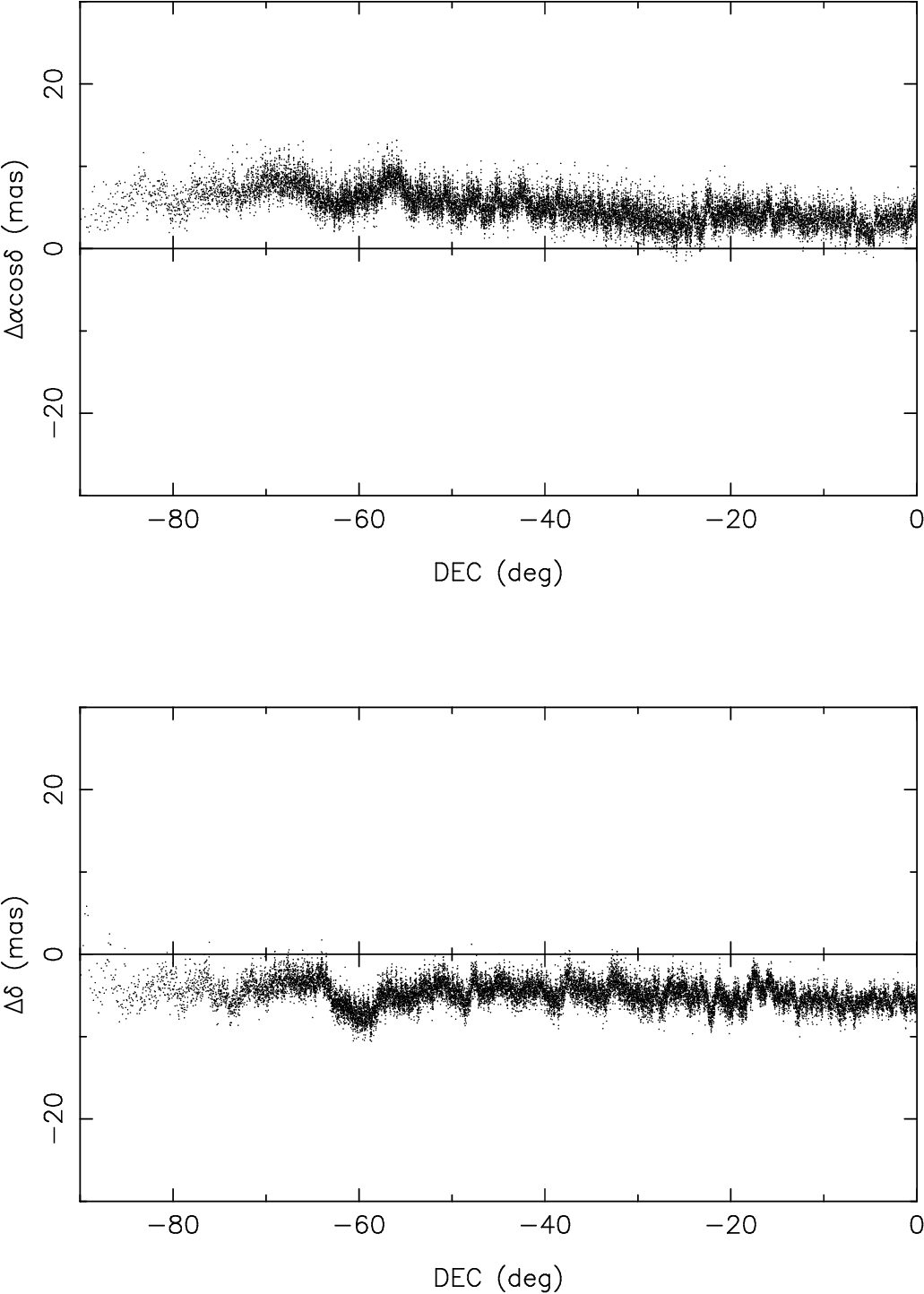}
\caption{Average position differences UCAC4$-$UCAC2 for
         southern hemisphere as a function of declination.}
\end{figure}

\begin{figure}
\epsscale{1.00}
\includegraphics[angle=0,scale=.42]{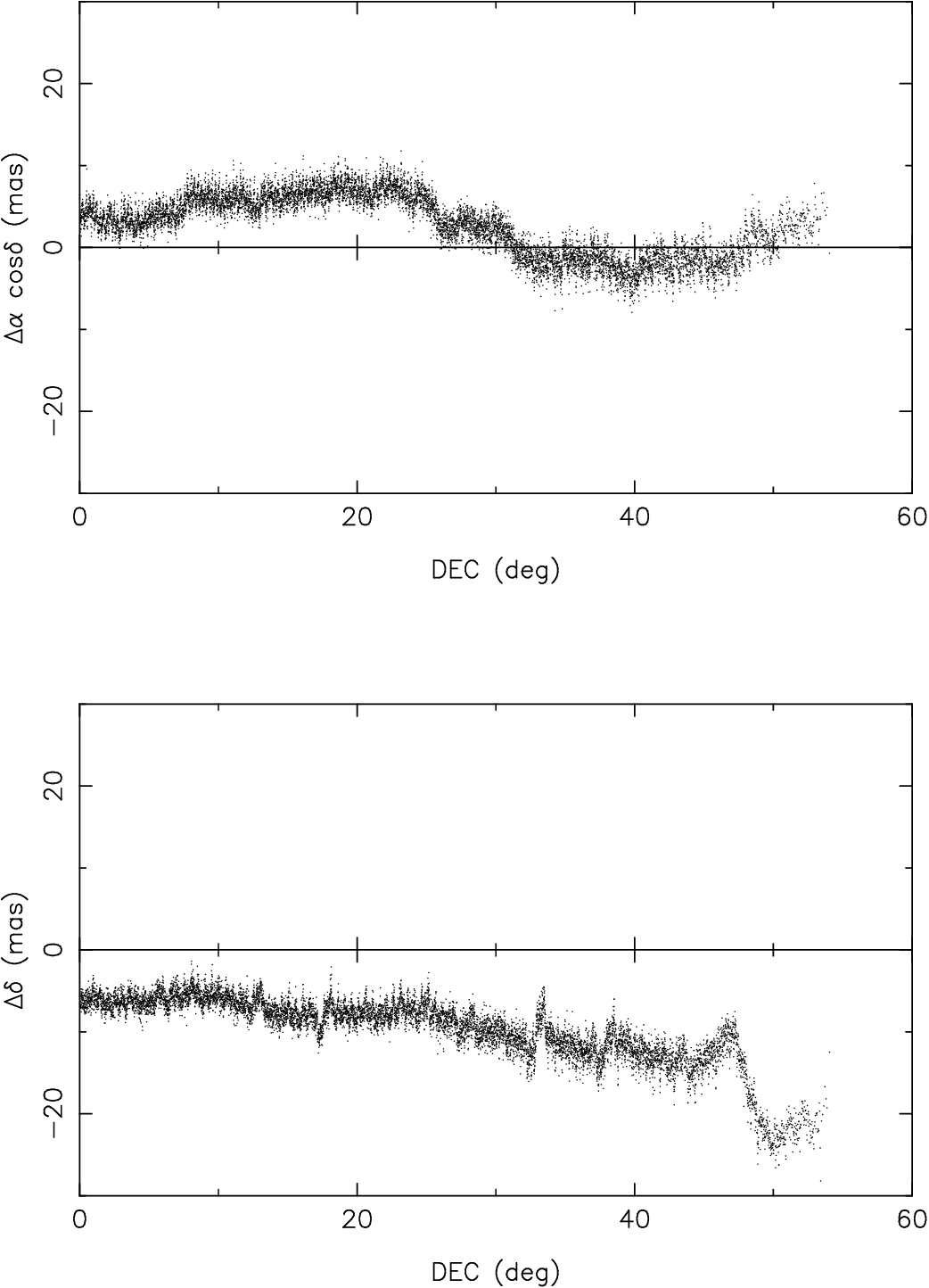}
\caption{The same as the previous figure for northern hemisphere data.}
\end{figure}

\clearpage

\begin{figure}
\epsscale{1.00}
\includegraphics[angle=0,scale=.45]{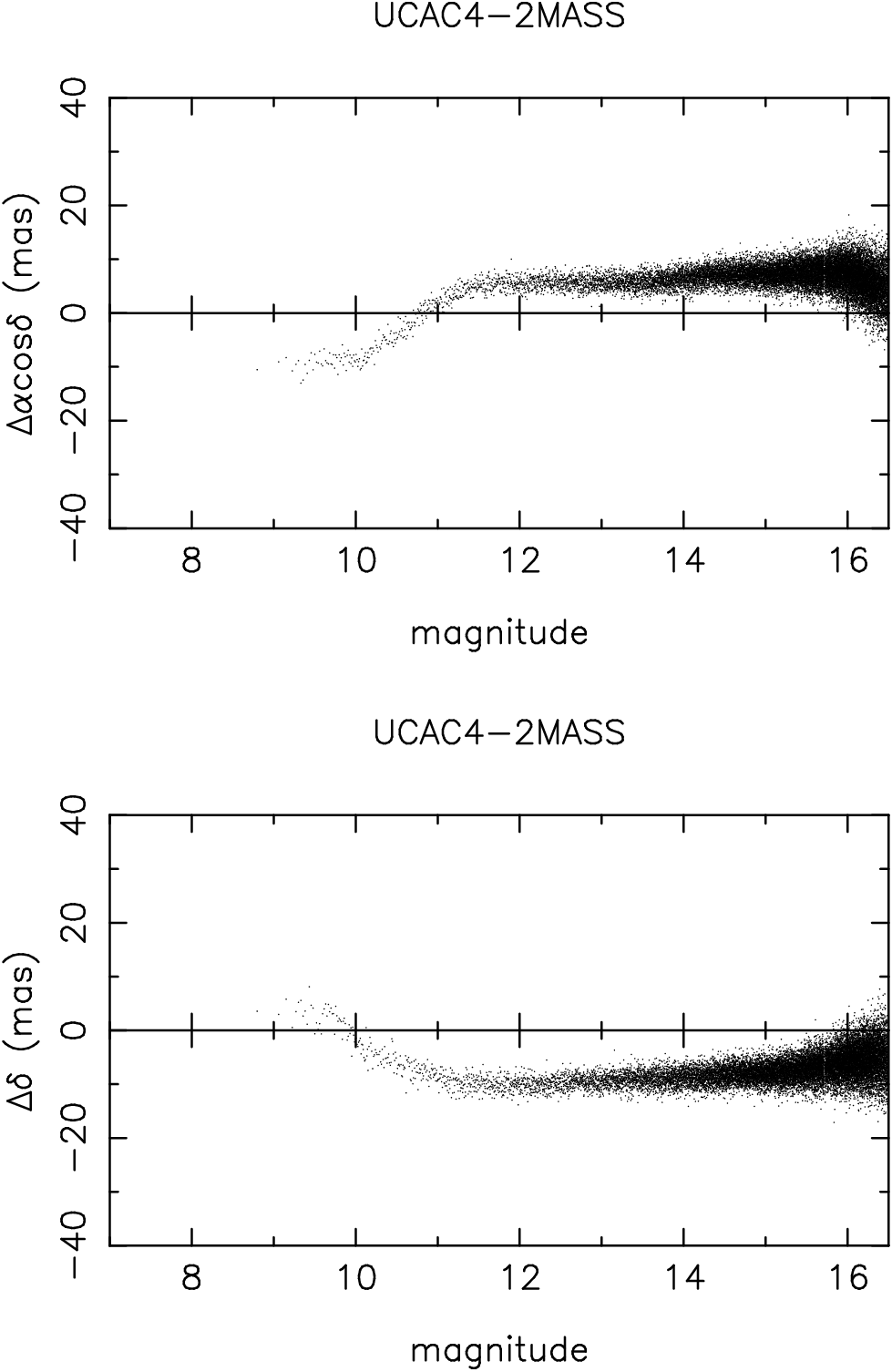}
\caption{Differences in position at epoch of 2MASS between UCAC4 and 
         2MASS (using UCAC4 proper motions) as a function of UCAC4
         aperture magnitude for southern declinations.}
\end{figure}

\begin{figure}
\epsscale{1.00}
\includegraphics[angle=0,scale=.45]{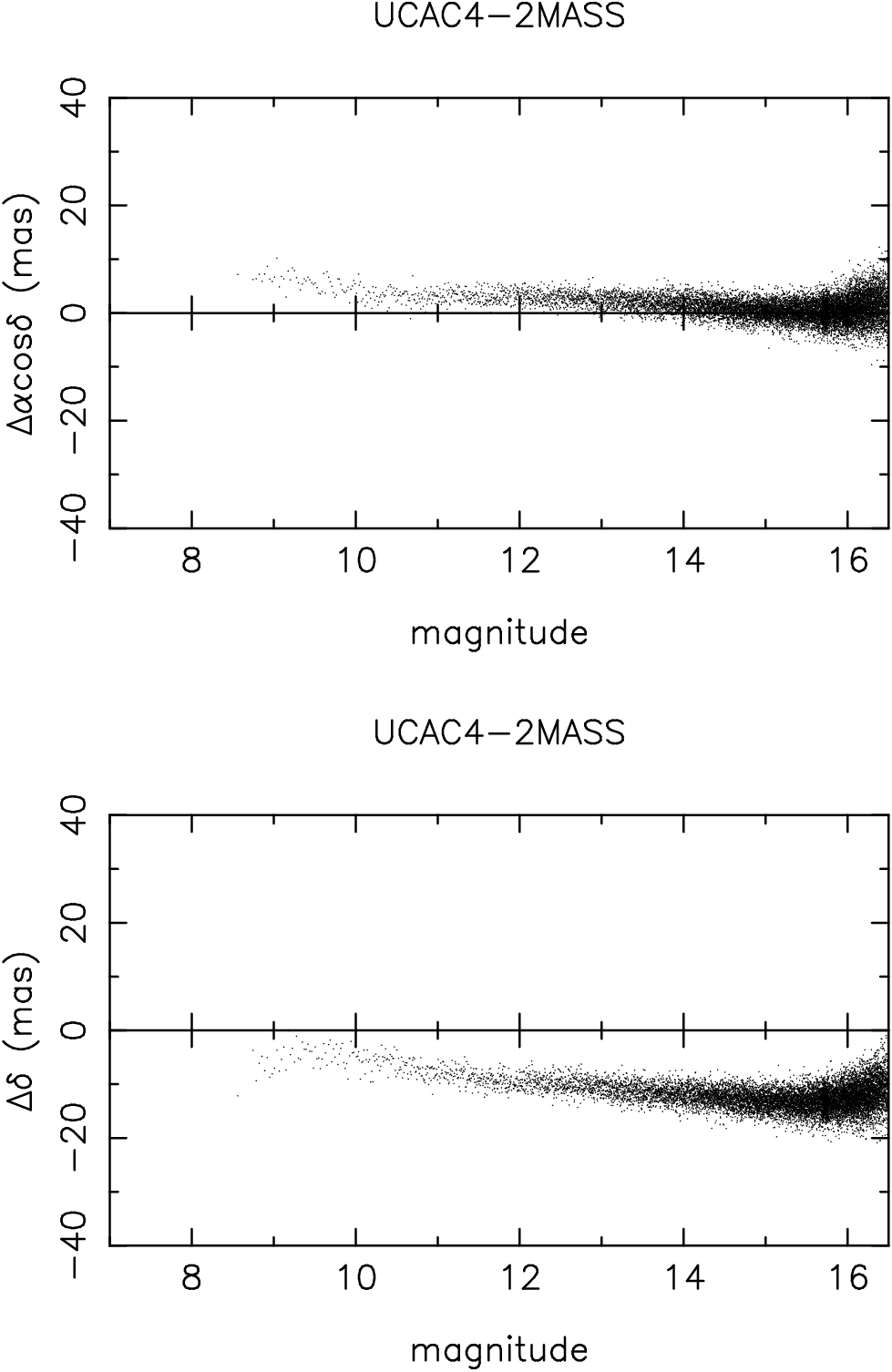}
\caption{The same as the previous figure for northern declinations.}
\end{figure}

\begin{figure}
\epsscale{1.00}
\includegraphics[angle=0,scale=.45]{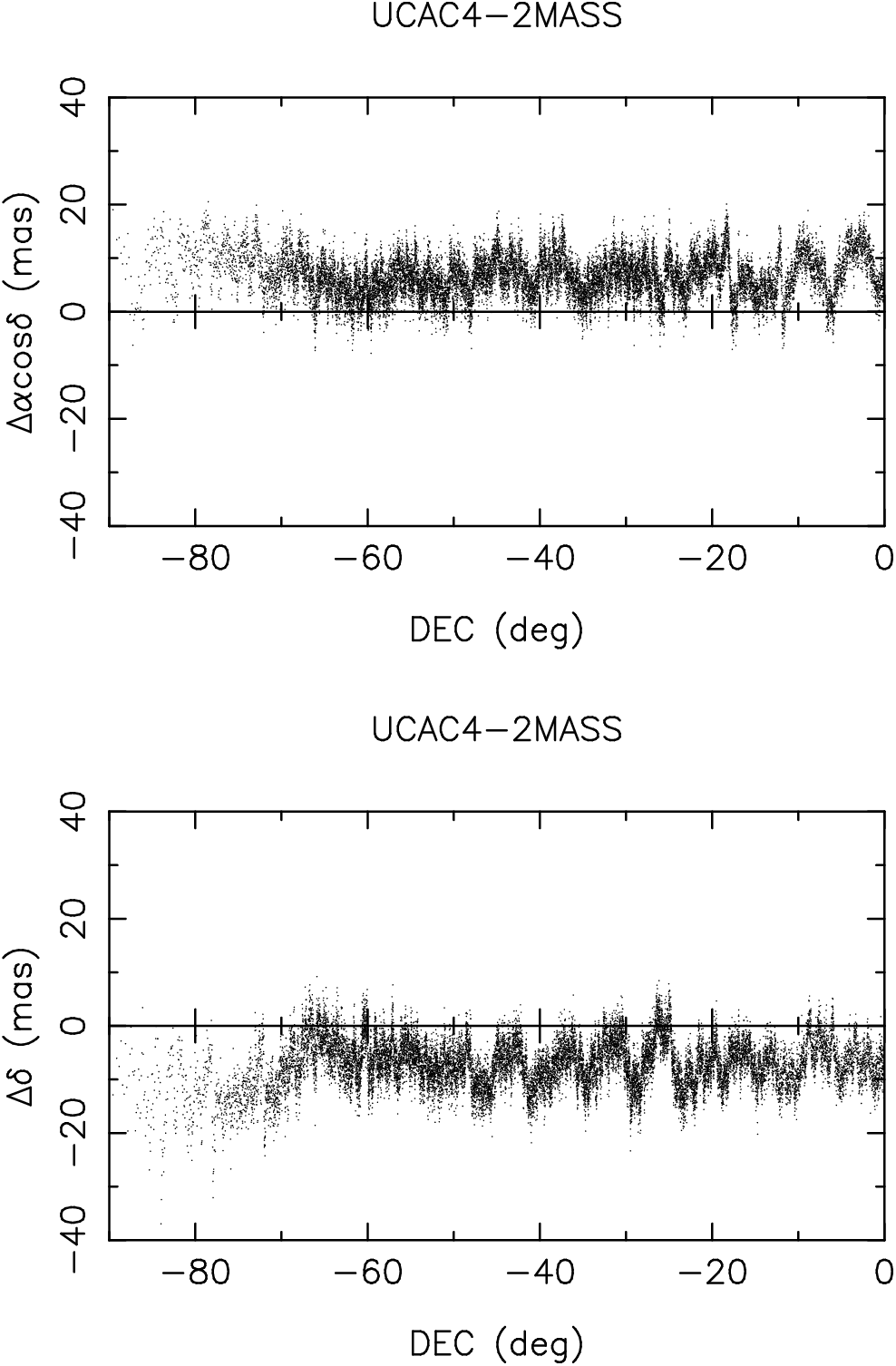}
\caption{Differences in position at epoch of 2MASS between UCAC4 and 
         2MASS (using UCAC4 proper motions) as a function of declination
         for southern declinations.}
\end{figure}

\begin{figure}
\epsscale{1.00}
\includegraphics[angle=0,scale=.45]{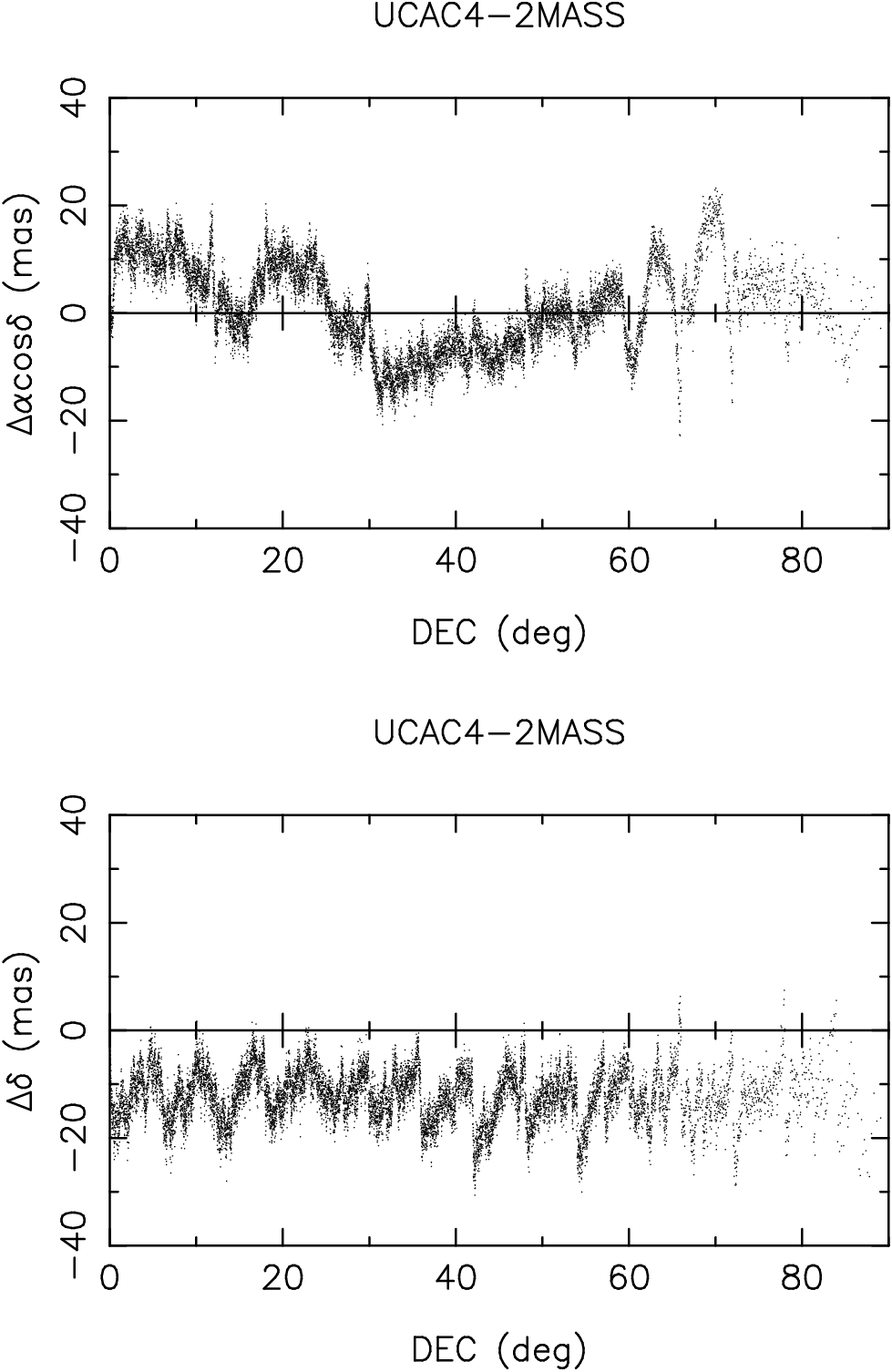}
\caption{The same as the previous figure for northern declinations.} 
\end{figure}

\begin{figure}
\epsscale{1.00}
\includegraphics[angle=0,scale=.45]{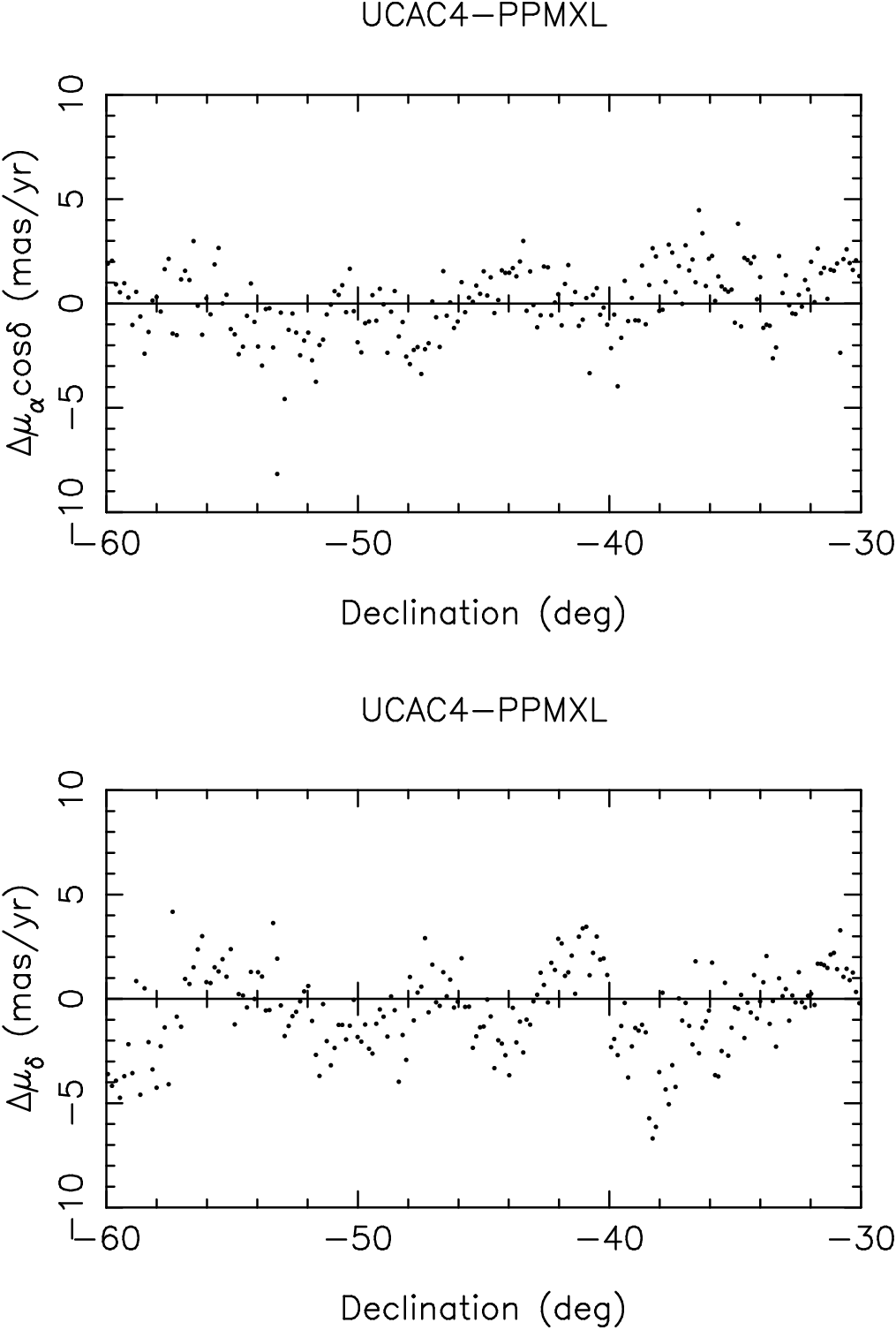}
\caption{Differences in proper motions (RA on top, Declination on
         bottom) between UCAC4 and PPMXL for stars in a narrow
         slice along RA = 6.0 to 6.1 hours as a function of Dec.
         One dot represents the mean over 200 stars.}
\end{figure}

\begin{figure}
\epsscale{1.00}
\includegraphics[angle=0,scale=.45]{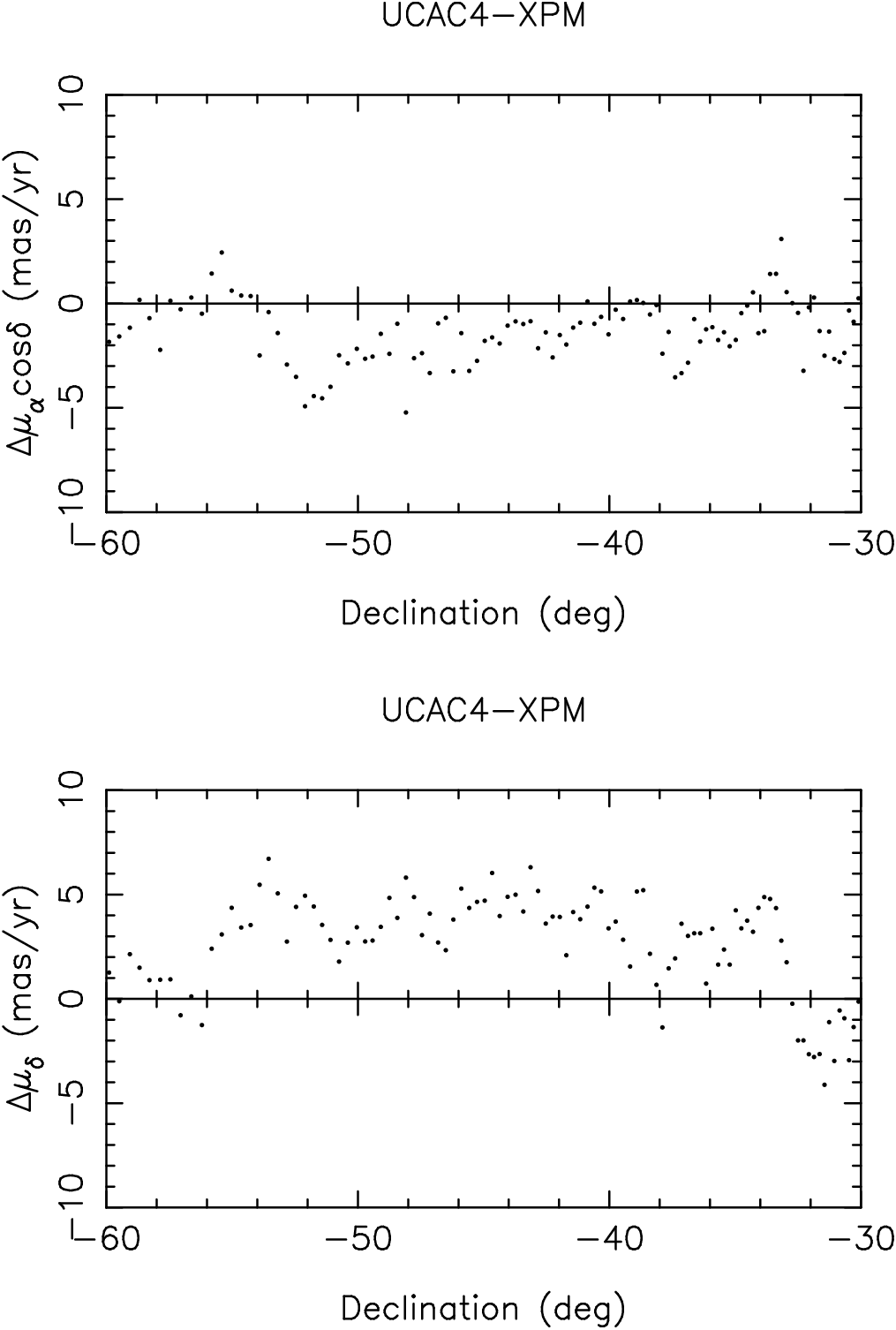}
\caption{The same as the previous figure for UCAC4$-$XPM proper
         motion differences.} 
\end{figure}

\begin{figure}
\epsscale{1.00}
\includegraphics[angle=0,scale=.45]{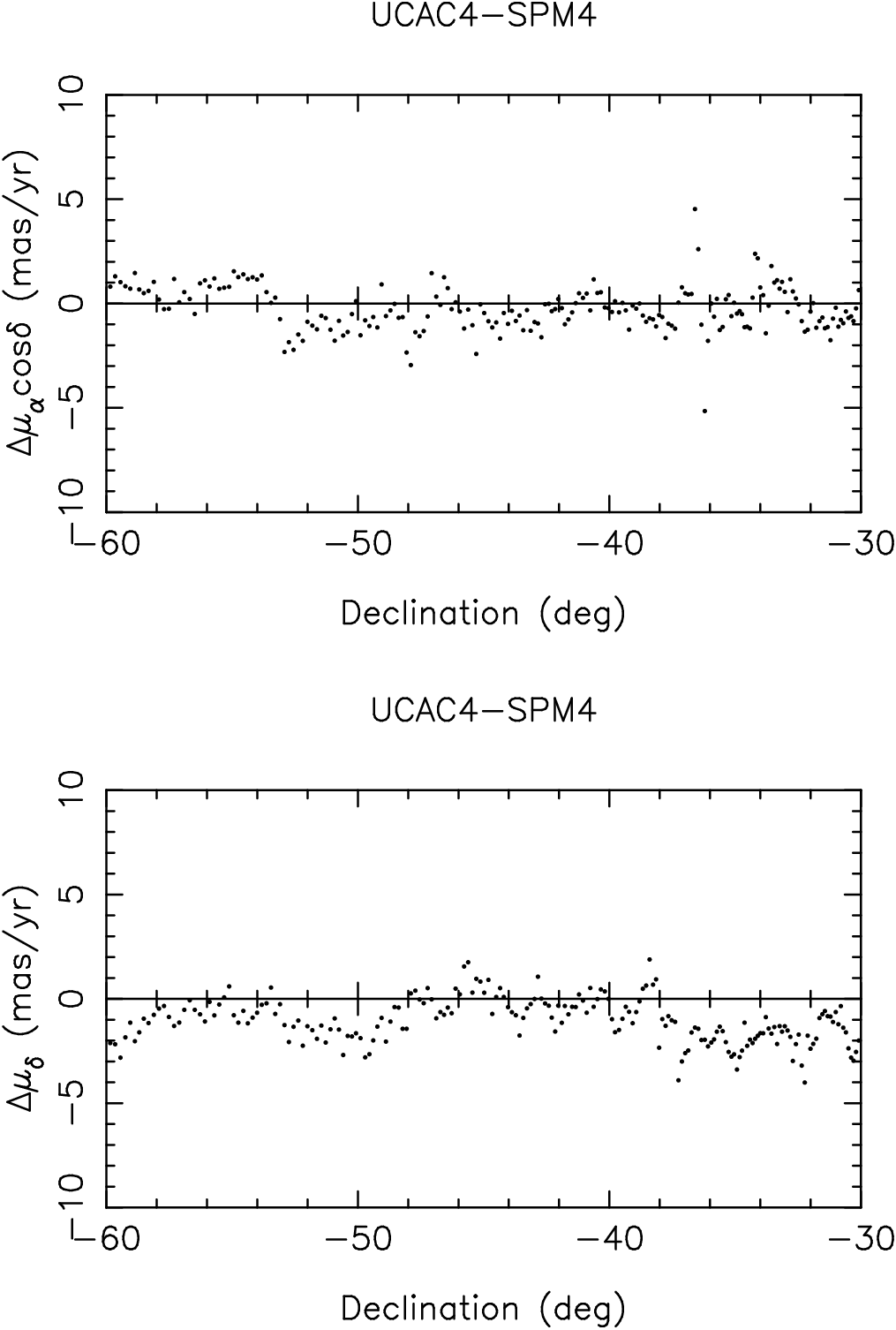}
\caption{The same as the previous 2 figures for UCAC4$-$SPM4
         proper motion differences.}
\end{figure}

\begin{figure}
\epsscale{1.00}
\includegraphics[angle=0,scale=.50]{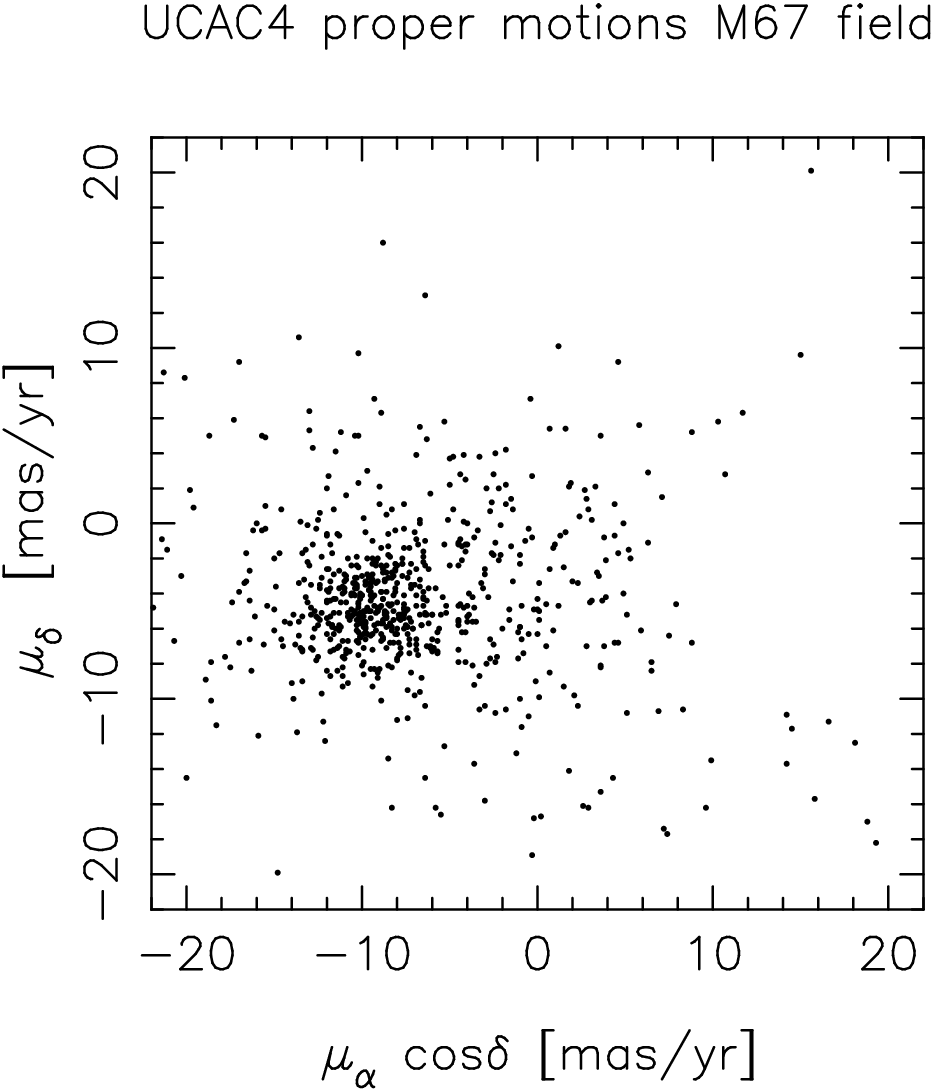}
\caption{UCAC4 proper motions RA are plotted versus proper
         motions Dec for a field centered on the open cluster M67.
         Stars are plotted within a radius of 20 arcmin from the
         center and only if their error in proper motion is
         $\le$ 7 mas/yr per component.}
\end{figure}

\clearpage


\begin{figure}
\epsscale{1.00}
\includegraphics[angle=-90,scale=.35]{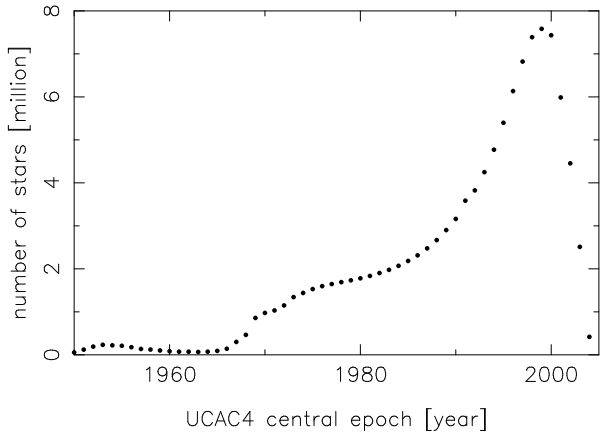}
\caption{Distribution of UCAC4 mean epochs.}
\end{figure}

\begin{figure}
\epsscale{1.00}
\includegraphics[angle=-90,scale=.35]{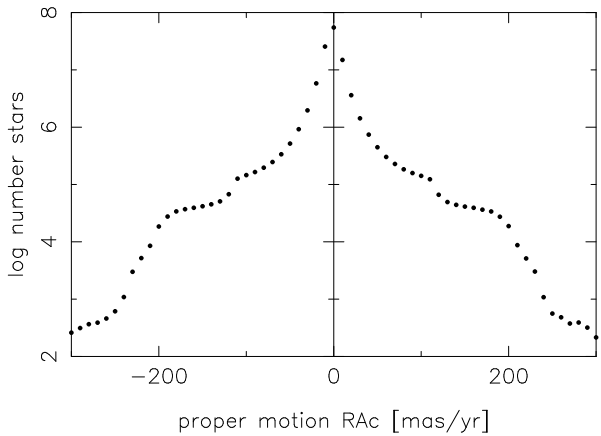}
\caption{Log of distribution of UCAC4 proper motions along RA.
         The plot of proper motions along Dec looks similar.}
\end{figure}

\begin{figure}
\epsscale{1.00}
\includegraphics[angle=-90,scale=.35]{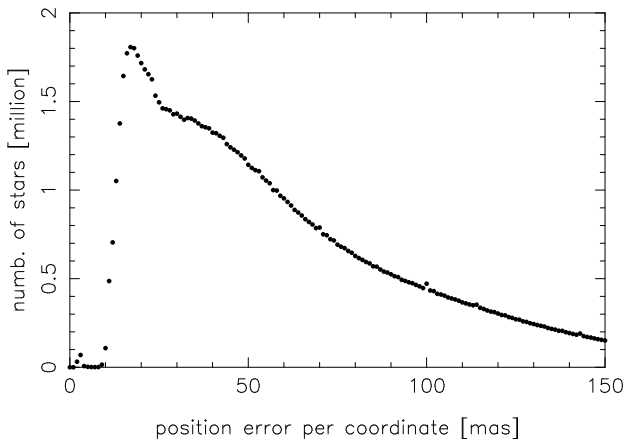}
\caption{Distribution of errors per coordinate for UCAC4 positions
   at epoch 2000, which is close to the mean epoch.}
\end{figure}

\begin{figure}
\epsscale{1.00}
\includegraphics[angle=-90,scale=.35]{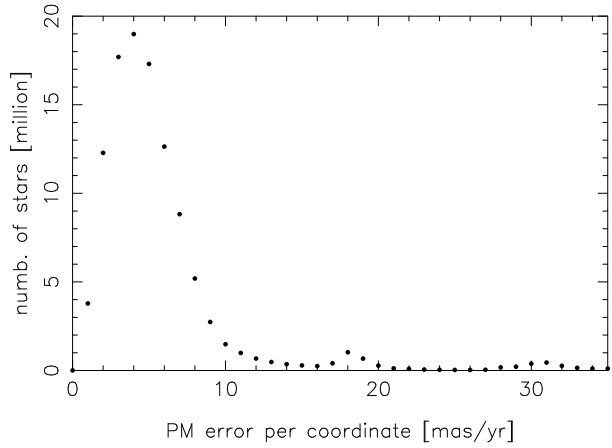}
\caption{Distribution of UCAC4 proper motion errors per coordinate.}
\end{figure}

\begin{figure}
\epsscale{1.00}
\plotone{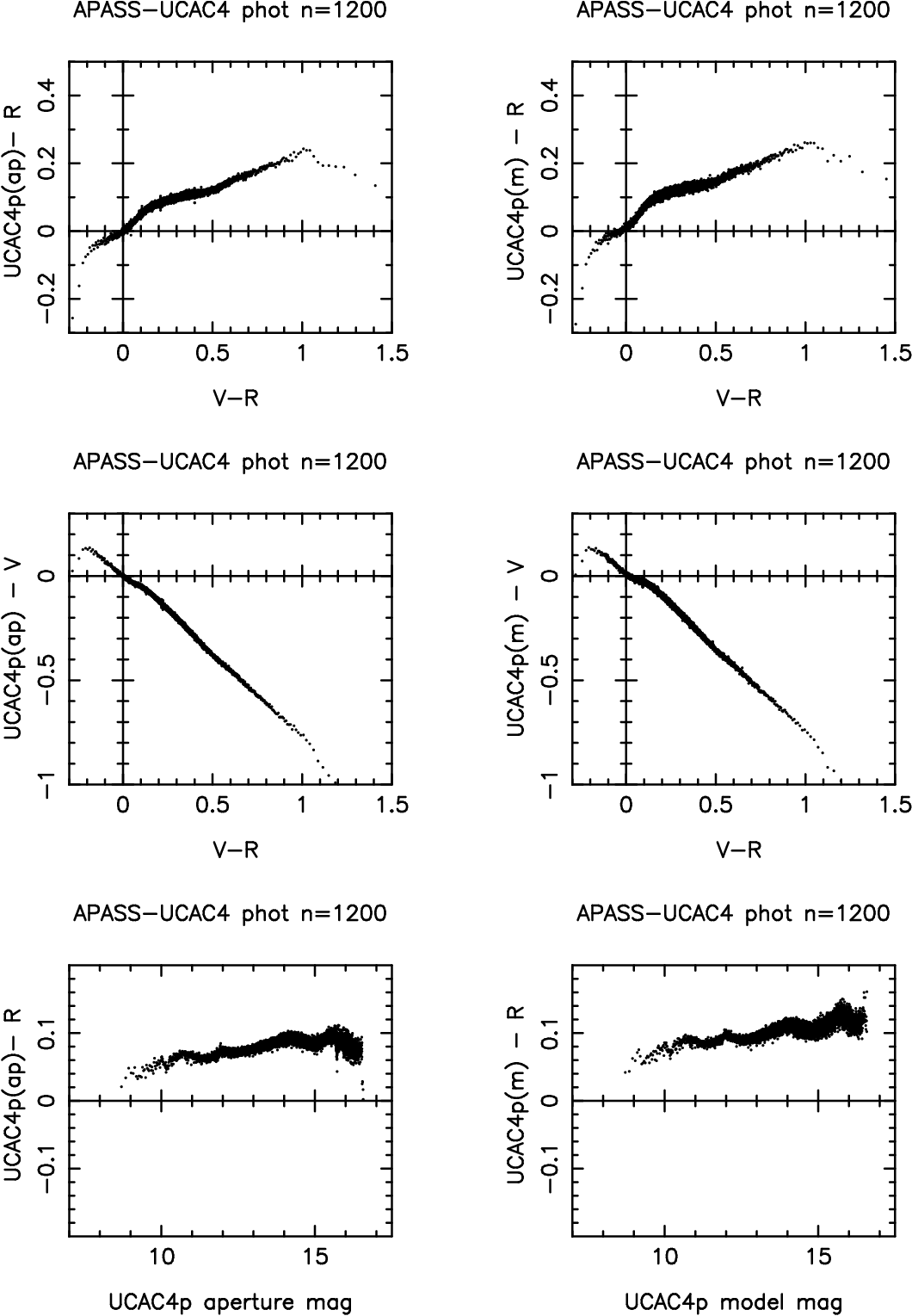}
\caption{Aperture (ap) and model (m) magnitudes of UCAC4p photometric
  system as calibrated with APASS V and r magnitudes.}
\end{figure}





\clearpage

\begin{table}
\begin{center}
\caption{General statistics of UCAC4 catalog entries.}
\vspace*{5mm}
\begin{tabular}{rl}  
\tableline\tableline
 number  & description  \\
of stars &    \\
\tableline
  113,780,093 & total number of entries (mostly stars) in UCAC4\\
  109,921,682 & with 2MASS identification \\
  106,689,821 & with proper motions \\
   81,897,551 & with 2 epoch proper motions \\
   27,245,403 & with 3 or more epoch proper motions \\
   80,806,744 & with 2 or more images from ``good fit" CCD obs.\\
   48,323,349 & matched with UCAC2 \\
       54,690 & matched with LEDA galaxies \\
       76,020 & matched with 2MASS extended source catalog \\
        8,925 & supplemented stars (no CCD observation) \\
      121,350 & with a matched Hipparcos star ID \\
      104,681 & with astrometry substituted by FK6/Hip/Tycho-2\\
\tableline
\end{tabular}
\end{center}
\end{table}


\begin{table}
\begin{center}
\caption{Number of stars in UCAC4 using other catalogs for proper motions.}
\vspace*{5mm}
\begin{tabular}{rl}  
\tableline\tableline
 number  &  catalog or data set name \\
of stars &    \\
\tableline
      120,487 & Hipparcos \\
    2,506,683 & Tycho-2 \\
    4,373,790 & AC2000 \\
      279,570 & AGK2 Bonn \\
      982,815 & AGK2 Hamburg \\
    4,682,287 & Hamburg Zone Astrograph \\
    3,492,601 & USNO Black Birch Astrograph, Yellow lens\\
    1,104,138 & Lick Observatory 50cm Astrograph \\
   68,887,550 & NPM (Lck1) \\
   57,355,612 & SPM Yale/San-Juan catalog (YSJ1) \\
\tableline
\end{tabular}
\end{center}
\end{table}


\begin{deluxetable}{rlcllr}
\tabletypesize{\scriptsize}
\tablecaption{Data items for each entry (star) in UCAC4}
\tablewidth{0pt}
\tablehead{
\colhead{item} & \colhead{label} & \colhead{format}\tablenotemark{a} & 
\colhead{unit} & \colhead{description}\tablenotemark{b} 
}
\startdata
 1& ra    &I*4&mas       & right ascension at  epoch J2000.0 (ICRS)\\
 2& spd   &I*4&mas       & south pole distance epoch J2000.0 (ICRS)\\
 3& magm  &I*2&millimag  & UCAC fit model magnitude                \\
 4& maga  &I*2&millimag  & UCAC aperture  magnitude                \\
 5& sigmag&I*2&0.01 mag  & UCAC error on magnitude                 \\
 6& objt  &I*1&          & object type                             \\
 7& cdf   &I*1&          & combined double star flag               \\
 8& sigra &I*1&mas       & m.e. at mean epoch in RA (*cos Dec)     \\
 9& sigdc &I*1&mas       & m.e. at mean epoch in Dec               \\
10& na1   &I*1&          & total numb. of CCD images of this star  \\ 
11& nu1   &I*1&          & numb. of CCD images used for this star  \\
12& cu1   &I*1&          & numb. of catalogs (epochs) used for PM  \\
13& cepra &I*2&0.01 yr   & mean epoch for RA, minus 1900           \\
14& cepdc &I*2&0.01 yr   & mean epoch for Dec,minus 1900           \\
15& pmrac &I*4&0.1 mas/yr& proper motion in RA*cos(Dec)            \\
16& pmdc  &I*4&0.1 mas/yr& proper motion in Dec                    \\
17& sigpmr&I*1&0.1 mas/yr& m.e. of pmRA * cos(Dec)                 \\
18& sigpmd&I*1&0.1 mas/yr& m.e. of pmDec                           \\
19& pts\_key&I*4&        & 2MASS pts key star identifier           \\
20& j\_m  &I*2&millimag  & 2MASS J  magnitude                      \\
21& h\_m  &I*2&millimag  & 2MASS H  magnitude                      \\
22& k\_m  &I*2&millimag  & 2MASS Ks magnitude                      \\
23& icqflg&I*1&          & 2MASS cc\_flg*10 + ph\_qual.flag for J  \\
24&  (2)  &I*1&          & 2MASS cc\_flg*10 + ph\_qual.flag for H  \\
25&  (3)  &I*1&          & 2MASS cc\_flg*10 + ph\_qual.flag for Ks \\
26& e2mpho&I*1& 0.01 mag & m.e. 2MASS J  magnitude j\_msigcom      \\
27&  (2)  &I*1& 0.01 mag & m.e. 2MASS H  magnitude h\_msigcom      \\
28&  (3)  &I*1& 0.01 mag & m.e. 2MASS Ks magnitude k\_msigcom      \\
29& apasm &I*2&millimag  &  B magnitude from APASS                 \\
30&  (2)  &I*2&millimag  &  V magnitude from APASS                 \\
31&  (3)  &I*2&millimag  &  g magnitude from APASS                 \\
32&  (4)  &I*2&millimag  &  r magnitude from APASS                 \\
33&  (5)  &I*2&millimag  &  i magnitude from APASS                 \\
34& apase &I*1& 0.01 mag & m.e. of B magnitude from APASS          \\
35&  (2)  &I*1& 0.01 mag & m.e. of V magnitude from APASS          \\
36&  (3)  &I*1& 0.01 mag & m.e. of g magnitude from APASS          \\
37&  (4)  &I*1& 0.01 mag & m.e. of r magnitude from APASS          \\
38&  (5)  &I*1& 0.01 mag & m.e. of i magnitude from APASS          \\
39& gcflg &I*1&          & Yale SPM g-flag*10 + c-flag             \\
40& icf   &I*4& merged   & FK6-Hipparcos-Tycho source flag         \\
41&  (2)  &...&          & AC2000       catalog match flag         \\
42&  (3)  &...&          & AGK2 Bonn    catalog match flag         \\
43&  (4)  &...&          & AGK2 Hamburg catalog match flag         \\
44&  (5)  &...&          & Zone Astrog. catalog match flag         \\
45&  (6)  &...&          & Black Birch  catalog match flag         \\
46&  (7)  &...&          & Lick Astrog. catalog match flag         \\
47&  (8)  &...&          & NPM Lick1    catalog match flag         \\
48&  (9)  &...&          & SPM YSJ1     catalog match flag         \\

49& leda  &I*1&          & LEDA galaxy flag                        \\
50& x2m   &I*1&          & 2MASS extend.source flag                \\
51& rnm   &I*4&          & unique star identification number       \\
52& zn2   &I*2&          & zone number of UCAC2 (0 = no match)     \\
53& rn2   &I*4&          & running record number along UCAC2 zone  \\
\enddata
\tablenotetext{a}{ ``I" means integer, followed by the number of bytes.}
\tablenotetext{b}{  ``m.e." stands for mean (or standard) error.}
\tablecomments{Extensive remarks are given only in the readme file of 
  the UCAC4 as part of the release data. This table strictly describes
  the data of the DVD release.  The online version served at CDS
  provides real numbers in ASCII format and declination instead of
  south pole distance.}
\end{deluxetable}


\begin{thebibliography}{}
\bibitem[Bellini et al. 2010]{Bel2010}
  Bellini, A., Bedin, L.R., Pichardo, B., Moreno, E., Allen, C.,
  Piotto, G., Anderson, J. 2010, A\&A, 513, 51
\bibitem[Boyd et al. 2011]{Boyd2011}
  Boyd, M.R., Winters, J.G., Henry, T.J., et al. 2011, \aj, 142, 10
\bibitem[Deacon et al. 2005]{Deacon2005}
  Deacon, N.R., Hambly, N.C., Cooke, J.A. 2005, A\&A, 435, 363
\bibitem[ESA 1997]{hipcat}
  The Hipparcos and Tycho Catalogues, European Space Agency, 1997,
  publication SP-1200
\bibitem[Fedorov et al. 2009]{xpm}
  Fedorov, P.N., Myznikov, A.A., Akhmetov, V.S. 2009, MNRAS 393, 133 
\bibitem[Finch et al. 2007]{Finch2007}
  Finch, C.T., Henry, T.J., Subasavage, J.P., Jao, W.C.,
  Hambly, N.C.  2007,  \aj, 133, 2898
\bibitem[Finch, Zacharias \& Wycoff 2010]{ucac3a}
  Finch, C., Zacharias, N. \& Wycoff, G.L. 2010, AJ 139, 2200
\bibitem[Finch, Zacharias \& Henry 2010]{uhpm1}
  Finch, C.T., Zacharias, N., Henry, T.J. 2010, \aj, 140, 844 
\bibitem[Finch et al.~2012]{uhpm2}
  Finch, C.T., Zacharias, N., Boyd, M.R., Henry, T.J., 
  Hambly, N.C. 2012, \apj, 745, 118 
\bibitem[Girard et al.~1998]{spm2}
  Girard,T. M., Platais,I., Kozhurina-Platais,V., van Altena, W. F.,
  Lopez, C. E. 1998, \aj, 115, 855 \\
   http://www.astro.yale.edu/spm/spm2cat/spm2.html
\bibitem[Girard et al.~2011]{spm4}
  Girard,T.M., van Altena,W.F., Zacharias,N., Viera,K., 
  Casetti-Dinescu,D.I., Castillo,D., Herrera,D., Lee,Y.S., 
  Beers,T.C., Monet,D.G., Lopez,C.E. 2011, AJ 142, 15
\bibitem[Henry et al. 2004]{Henry2004}
  Henry, T.J., Subasavage, J.P., Brown, M.A., Beaulieu, T.D.,
  Jao, W.C., Hambly, N.C.  2004,
  \aj, 128, 2460
\bibitem[H{\o}g et al.~2000]{tycho2}
   H{\o}g, E. Fabricius, C., Makarov, V., Urban, S.,
   Corbin, T., Wycoff, G., Bastian, U., Schwedendiek, P.,
   Wicenec, A. 2000, A\&A, 355, L27
\bibitem[Hanson, et al. 2004]{npm2}
  Hanson, Klemola, Jones, and Monet 2004; AJ,128,1430
\bibitem[Klemola, Jones, Hanson 1987]{npm1}
  Klemola,A., Hanson, Jones, 1987, AJ 94, 501
\bibitem[Lepine 2008]{Lep2008}
   Lepine, S.  2008, \aj, 135, 2177
\bibitem[Lepine \& Shara 2005]{LS2005}
   Lepine,S., Shara,M.M. 2005, \aj, 129, 1483
\bibitem[Monet \& Levine 2001]{pmm}
   Monet, D. G. \& Levine, S. E. 2001, ASP Conf.~232, 284,
   Eds. R.~Clowes, A.~Adamson, \& G.~Bromage, San Francisco
\bibitem[Roeser et al. 2010]{ppmxl}
   Roeser, S., Demleitner, M., Schilbach, E. 2010, \aj 139, 2440 
\bibitem[Ruiz et al. 2001]{Ruiz2001}
   Ruiz, M.T., Wischnjewsky, M.R., Patricio, M., Gonzalez, L.E. 2001,
   ApJ, 133, 199
\bibitem[Salim \& Gould 2003]{SG2003}
   Salim S., Gould A. 2003, \aj, 582, 1011
\bibitem[Scholz et al. 2000]{Scholz2000}
   Scholz,R.-D., Irwin,M., Ibata,R., Jahreiß,H., Malkov,O. Yu. 2000,
   A\&A, 353, 958
\bibitem[Skrutskie, Cutri, Stiening et al.~2006]{2mass}
  Skrutskie, M. F, Cutri, R. M., Stiening, R. et al. 2006,
  \aj, 131, 1163
\bibitem[Subasavage et al. 2005a]{Suba}
  Subasavage, J.P., Henry, T.J., Hambly, N.C., Brown, M.A.,
  Jao, W.C. 2005, \aj, 129, 413
\bibitem[Subasavage et al. 2005b]{Subb}
  Subasavage, J.P., Henry, T.J., Hambly, N.C., Brown, M.A.,
  Jao, W.C., Finch, C.T. 2005, \aj, 130, 1658
\bibitem[Urban et al.~2000]{ac2000}
  Urban, S.E., Corbin, T.E., Wycoff, G.L., Makarov, V.V., Hoeg, E.,
     Fabricius, C.  2000, BAAS, 33, 1494
\bibitem[van Leeuwen 2007]{hip2}
   van Leeuwen, F. 2007, Springer Science Library, Vol. 350
\bibitem[Wielen et al. 1999]{fk6}
  Wielen,R., Schwan,H., Dettbarn,C., Lenhardt,H. Jahreiss,H.,
  Jahrling,R. 1999, Veroeff. Astron. Rechen-Inst. Heidelberg 35, 1
\bibitem[Wroblewski \& Costa 1999]{WC1999}
  Wroblewski, H. \& Costa, E.  1999, A\&A, 139, 25 
\bibitem[Wroblewski \& Costa 2001]{WC2001}
  Wroblewski, H. \& Costa, E.  2001, A\&A, 367, 725
\bibitem[Wroblewski \& Torres 1989]{WT1989}
  Wroblewski, H. \& Torres, C. 1989, A\&A, 78, 231
\bibitem[Wroblewski \& Torres 1991]{WT1991}
  Wroblewski, H. \& Torres, C. 1991, A\&A, 91, 129
\bibitem[Wroblewski \& Torres 1994]{WT1994}
  Wroblewski, H. \& Torres, C. 1994, A\&A, 105, 179
\bibitem[Wroblewski \& Torres 1996]{WT1996}
  Wroblewski, H. \& and Torres, C. 1996, A\&A, 115, 481
\bibitem[Wroblewski \& Torres 1997]{WT1997}
  Wroblewski, H. \& Torres, C. 1997, A\&A, 122, 447
\bibitem[Zacharias, Zacharias \& Finch 2012]{mizIAU12}
  Zacharias, M.I., Zacharias,N., Finch,C. 2012,
  IAU GA, Bejing, Joint Discussion 7 (in prep.)
\bibitem[Zacharias 1992]{BAnz}
  Zacharias, N., 1992, A\&A 264, 397
\bibitem[Zacharias et al.~2000]{ucac1}
  Zacharias, N., Zacharias, M. I., \& Rafferty, T. J. 2000, \aj, 118, 2503
\bibitem[Zacharias et al.~2004]{ucac2}
  Zacharias, N., Urban, S. E., Zacharias, M. I., Wycoff, G. L.,
  Hall, D. M., Monet, D. G., \& Rafferty, T. J. 2004, \aj, 127, 3043
\bibitem[Zacharias et al.~2008]{starscan}
  Zacharias, N., Winter, L, Holdenried, E.R., De Cuyper,J.-P.,
  Rafferty, T.J., Wycoff, G.L.   2008,
   \pasp, 120, 644 
\bibitem[Zacharias 2010]{ucac3x}
  Zacharias, N. 2010, \aj, 139, 2208
\bibitem[Zacharias et al. 2010]{ucac3r}
  Zacharias, N., Finch, C., Girard, T., Hambly, N., Wycoff, G.,
  Zacharias, M.I., et al. 2010, \aj, 139, 2184
\end{thebibliography}
\end{document}